\documentclass{article}

\usepackage{PRIMEarxiv}

\usepackage[utf8]{inputenc} 
\usepackage[T1]{fontenc}    
\usepackage{hyperref}       
\usepackage{url}            
\usepackage{booktabs}       
\usepackage{amsfonts}       
\usepackage{nicefrac}       
\usepackage{microtype}      
\usepackage{lipsum}
\usepackage{fancyhdr}       
\usepackage{graphicx}       
\usepackage{amsmath}
\usepackage{amsfonts}
\usepackage{array}
\usepackage{mdwmath}
\usepackage{mdwtab}
\usepackage{eqparbox}
\usepackage{enumitem}
\usepackage{url}
\usepackage{cite}
\usepackage{stfloats}
\graphicspath{{media/}}     

\title{Electromagnetic Normalization of Channel Matrix for Holographic MIMO Communications}

\author{Shuai S. A. Yuan $^{1}$, Li Wei $^{2}$, Xiaoming Chen $^{3}$, Chongwen Huang $^{1}$, and Wei E. I. Sha $^{1,}$*\\ \\
	$^{1}$ College of Information Science and Electronic Engineering, Zhejiang University, Hangzhou 310027, China.\\
	$^{2}$ School of Electrical and Electronics Engineering, Nanyang Technological University, Singapore 639798.\\
	$^{3}$ School of Information and Communications Engineering, Xi'an Jiaotong University, Xi'an 710049, China.\\
\\ \\
	\texttt{Correspondence: weisha@zju.edu.cn}}

\begin{document}
\maketitle

\begin{abstract} %
	Holographic multiple-input and multiple-output (MIMO) communications introduce innovative antenna array configurations, such as dense arrays and volumetric arrays, which offer notable advantages over conventional planar arrays with half-wavelength element spacing. However, accurately assessing the performance of these new holographic MIMO systems necessitates careful consideration of channel matrix normalization, as it is influenced by array gain, which, in turn, depends on the array topology. Traditional normalization methods may be insufficient for assessing these advanced array topologies, potentially resulting in misleading or inaccurate evaluations. In this study, we propose electromagnetic normalization approaches for the channel matrix that accommodate arbitrary array topologies, drawing on the array gains from analytical, physical, and full-wave methods. Additionally, we introduce a normalization method for near-field MIMO channels based on a rigorous dyadic Green's function approach, which accounts for potential losses of gain at near field. Finally, we perform capacity analyses under quasi-static, ergodic, and near-field conditions, through adopting the proposed normalization techniques. Our findings indicate that channel matrix normalization should reflect the realized gains of the antenna array in target directions. Failing to accurately normalize the channel matrix can result in errors when evaluating the performance limits and benefits of unconventional holographic array topologies, potentially compromising the optimal design of holographic MIMO systems.
\end{abstract}




\section{Introduction}
Based on Shannon's information theory \cite{shannon}, multiple-input-multiple-output (MIMO) technology utilizing spatial multiplexing has been developed to enhance the channel capacity of modern wireless communications \cite{telatar1999capacity, TL2014}. To further improve MIMO performance, massive antennas are employed \cite{bjornson2019massive}, leading to either holographic MIMO with small element spacing and fixed array areas \cite{Huang2020,Pizzo2020} or extremely large antenna arrays (ELAA) with larger element spacing and total array area \cite{lu2021communicating, wang2024tutorial}. Given that the array size at base stations is often constrained due to factors such as wind drag, assembly complexity, weight, and urban planning regulations, the development of holographic MIMO with fixed array areas is particularly important for current base-station applications. Approaching a near-continuous aperture, holographic MIMO enables unprecedented electromagnetic wave manipulation, such as exploiting evanescent waves \cite{wei2023tri} and achieving super-directivity \cite{ji2024electromagnetic}, showing potential in enhancing spectral efficiency, power transfer, sensing, and so on \cite{gong2023holographic, Dardari2021,An2023}.

Although holographic MIMO demonstrates advantages in many scenarios, certain limitations persist \cite{franceschetti_2017}. The benefits of MIMO communications generally fall into two categories: degrees of freedom (DOF) gain and beamforming gain, both of which are inherently limited by the array size. When the element spacing is reduced to half a wavelength, performance gains tend to plateau. However, holographic arrays need not be confined to planar configurations. Volumetric arrays, or three-dimensional (3-D) arrays, have been proposed to overcome these limitations \cite{yuan2024breaking}. Physically, volumetric arrays can achieve larger effective or shadow area along different radiation directions \cite{gustafsson2024shadow}, leading to improved array gains and DOF. Given these observed benefits, the exploration of diverse regular and irregular volumetric array topologies is expected to become an increasingly attractive area of research.

When designing and evaluating novel holographic array topologies, careful attention must be given to the normalization of the channel matrix. While DOF gain is determined by the rank of the channel matrix, beamforming gain is associated with its Frobenius norm. The evaluation of DOF is unaffected by the normalization of the channel matrix, but capacity is influenced by the channel normalization, as it directly impacts the scaling of the signal-to-noise ratio (SNR). A common approach is to use unit sub-channel gain for normalization by making $||\mathbf{H}||_F^2=N_tN_r$, where $||\mathbf{H}||_F$ is the Frobenius norm of channel matrix $\mathbf{H}$, $N_t$ and $N_r$ are the transmitting and receiving antenna numbers. However, this method depends only on the number of antennas and not on the array topology, making it only approximately valid for linear or planar arrays with near half-wavelength spacing, but not applicable for dense or volumetric arrays. The Frobenius norm of the channel matrix can be physically connected to the product of array gains at the transmitter and receiver sides. In holographic arrays, where antenna elements are densely packed, the correlations between antennas cannot be ignored. Consequently, electromagnetic factors and constraints associated with holographic array gain, i.e., array topology, must be considered. Failing to do so may result in an inaccurate scaling of SNR, leading to misleading capacity outcomes. These inaccuracies could manifest as incorrect trends in performance evaluation or misestimated benefits of new array topologies, which is critical when advancing novel holographic MIMO systems.

\subsection{Prior works}
Much research has been conducted on the normalization of the channel matrix. The most widely used method in massive MIMO involves channel normalization by setting unit sub-channel gain. This approach works effectively in both channel simulations and experiments, particularly when the element spacing is not too close and regular array topologies, such as uniform linear or planar arrays, are employed \cite{gao2015massive, wallace2003experimental, mabrouk2013feasibility}. In addition to normalizing the entire channel matrix, normalization can also be applied to each user separately, i.e., each column of the channel matrix. This method helps eliminate power imbalances caused by different path losses experienced by different users \cite{martinez2018experimental}. These two approaches can be referred to as matrix and vector normalizations, respectively. The impact of these normalization methods on massive MIMO precoding has also been investigated \cite{sadeghi2017large, Caire2015}. Beyond these conventional normalization methods, electromagnetic-based normalization approaches have been proposed \cite{Loyka2009}. These methods show that channel normalization is related to the antenna array gains on both the transmitter and receiver sides and depends on the spatial multiplexing strategy. Similar channel power gain constraints are also introduced in the analysis of holographic MIMO \cite{d2024holographic}, MIMO capacity bounds with finite scatterers \cite{burr2003capacity}, and electromagnetic information theory analysis \cite{zhu2024mimo}, as the capacity of MIMO systems would otherwise increase infinitely with the number of antennas if no physical constraints on power gain, i.e., channel normalization, are applied. Furthermore, normalization methods that account for cross-polarization have been proposed \cite{coldrey2008modeling}, which can address polarization discrimination in complex or potential near-field scenarios. Additionally, using normalization based on array gain formulae allows for the inclusion of antenna effects in the capacity analysis for line-of-sight (LOS) scenarios \cite{qian2024including}.

Previous studies on normalization have primarily concentrated on regular linear or planar arrays, often relying on analytical formulae to analyze dense arrays while considering power gain limits. However, these analytical formulae sometimes fall short in accurately capturing gain behaviors. For instance, the analytical formulae for linear arrays are limited by their ability to account for the decrease in gain at large $\theta$ angles because they cannot incorporate the concept of `area'—an aspect that can only be accurately represented through physical or full-wave simulation methods. Furthermore, as an emerging technique, the normalization of volumetric arrays has not yet been extensively investigated. In addition, the decrease in antenna efficiency, which is unavoidable in dense holographic arrays, is often not considered in these analyses. Potential losses in near-field gain due to factors such as cross-polarization and non-uniform illumination are also frequently overlooked. These neglected factors may significantly impact the design and evaluation of novel holographic arrays, potentially leading to inaccuracies in performance assessments and suboptimal array designs.

\subsection{Our contributions}
In this paper, we aim to incorporate the above-mentioned electromagnetic factors into existing normalization methods, facilitating electromagnetic-based normalization for arbitrary antenna arrays in holographic MIMO communications. The contributions of this paper can be summarized as follows:
\begin{itemize} 
	\item We present electromagnetic normalization approaches for the far-field MIMO channel matrix tailored to linear, planar, and volumetric arrays, grounded in the average far-field gains calculated through analytical, physical, and full-wave methods. Particularly, we reveal some properties of linear and planar arrays that traditional analytical methods fail to capture, and we provide a comprehensive discussion of the normalization for volumetric arrays—a research direction that has not been previously explored.
	\item We present a novel approach to the normalization of the near-field MIMO channel matrix, utilizing rigorous dyadic Green's function and Poynting vector methods. In doing so, we uncover that additional loss factors—such as cross-polarization, non-uniform illumination, and non-ideal beamforming—may be introduced beyond the far-field gain limits. To facilitate the practical application of this method, we also introduce a physical loss factor formula that aligns closely with our rigorous analyses.
	\item Capacity analyses in quasi-static, ergodic, and near-field scenarios are conducted using the proposed normalization methods. Numerical results demonstrate that inappropriate normalization can lead to inaccurate performance evaluations of unconventional array topologies. Additionally, the benefits of volumetric arrays are highlighted, showing advantages in both array gain and capacity.
	
\end{itemize}

The remainder of this paper is organized as follows. Section II establishes the problem statement for the normalization of holographic MIMO channel matrices. Section III explores the far-field gains of arbitrary array topologies using analytical, physical, and full-wave methods. In Section IV, the near-field gain of antenna arrays is examined through the dyadic Green's function and Poynting vector approaches. Section V presents capacity analyses in quasi-static, ergodic, and near-field scenarios, utilizing proposed normalization methods. Finally, Section VI concludes the paper with a summary of findings and implications.

Notation: Fonts $a$, $\mathbf{a}$, and $\mathbf{A}$ represent scalars, vectors, and matrices, respectively. $\mathbf{A}^{T}$, $\mathbf{A}^{\dagger}$, and $||\mathbf{A}||_{F}$ denote transpose, Hermitian (conjugate transpose), and Frobenius norm of $\mathbf{A}$, respectively. $^{*}$ is the conjugate operator. $\mathbf{a}_i$ denotes the $i$th column of $\mathbf{A}$, and $||\mathbf{a}_i||$ is the norm of $\mathbf{a}_i$. $\mathrm{Re}(\cdot)$ is the real part of the argument. $\mathbf{I}_n$ (with $n \leq 2$) is the $n\times n$ identity matrix. Notation $\mathbb{E} [\cdot] $ denotes the expectation of the argument. $a !$ is the factorial of $a$, and $(a)_k$ is the falling factorial of $a$ with order $k$. 

\section{Problem Statement}
\subsection{Quasi-static model}
Considering the standard baseband model of a quasi-static MIMO channel \cite{tse2005fundamentals}
\begin{equation}
\mathbf{y}=\mathbf{H x}+\mathbf{w},
\end{equation}
where $\mathbf{x}$ and $\mathbf{y}$ are the transmitter (Tx) and receiver (Rx) signal vectors, $\mathbf{w}$ is the noise vector, $\mathbf{H} \in \mathbb{C}_{N_t \times N_r}$ is the MIMO channel matrix. Adopting the conventional Bell Laboratories Layer Space-Time (BLAST) type spatial multiplexing strategy\cite{tse2005fundamentals}, i.e., Tx antennas are uncorrelated and equally-power allocated, we have the well-known capacity formula
\begin{equation}
C=\log \left[\operatorname{det}\left(\mathbf{I}_{N_t\times N_t}+\frac{\gamma}{N_t} \mathbf{H H}^{\dagger}\right)\right],
\end{equation}
where $\gamma$ is the total signal-to-noise ratio (SNR). All the physical characteristics of the MIMO system are encapsulated in the matrix $\mathbf{H}$: (a) The rank of $\mathbf{H}$ represents the communication DOF, indicating the number of parallel channels available for information transmission; (b) The Frobenius norm of $\mathbf{H}$ reflects the total channel gain, which arises from the beamforming gains of the Tx and Rx antenna arrays. The communication DOF has been extensively studied \cite{miller2019waves, ShuaiPra, bai2024information} and is independent of channel matrix normalization. However, the norm of $\mathbf{H}$ has not been well investigated, particularly for novel array topologies with beamforming gains that differ from conventional arrays. Traditionally, the channel matrix is normalized by making
\begin{equation}\label{eq3}
\|{\mathbf{H}}\|^2_F=N_t N_r,
\end{equation}
where unit sub-channel gain is assumed, and the antenna number $N_t$ and $N_r$ here in fact represent the power gains at Tx and Rx sides. Strictly speaking, using this normalization is only approximately valid for linear or planar array with half-wavelength element spacing \cite{Loyka2009}. However, in current holographic MIMO applications that utilize dense or even volumetric arrays, the physical insights into the gain introduced by these novel array topologies cannot be adequately captured with traditional normalization method, including limited effective area, coupling effects, and extra gain benefits from volumetric array. Generally, the appropriate normalization method for the channel matrix depends on the coherence at the transmitter and receiver sides, which can be categorized into the following three cases.
\subsubsection{Tx non-coherent, Rx coherent}
This is BLAST type MIMO communication, which is widely accepted in current 5G applications. In a typical urban scenario of 3rd Generation Partnership Project (3GPP) standard, the users (Tx side) are distributed in a certain area with low correlations between each other, thus are regarded as non-coherent. The Rx side at base station could be a holographic antenna array, where the correlations between antennas cannot be neglected due to close spacing. In this case, we have
\begin{equation}
P_r=\sum_{i=1}^{N_t}\left\|\mathbf{h}_i\right\|^2 \frac{P_t}{N_t}=\|\mathbf{H}\|^2_F \frac{P_t}{N_t},
\end{equation}
where $P_t$ and $P_r$  are the Tx and Rx power, and they fulfill the Friis transmission equation
\begin{equation}
\label{Friis}
P_r=P_t \frac{{G}_r {G}_t}{L_p},
\end{equation}
where $G_r$ and $G_t$ are the array gains at Rx and Tx sides. In order to focus on the effect of fast fading due to multi-path effect, the distance factor $L_p$ is eliminated by setting it as 1. As the transmitters here are uncorrelated and no constructive interference can be generated, $G_t$ is equal to 1 in this case. Therefore, the channel matrix should be normalized as
\begin{equation}\label{6}
\|{\mathbf{H}}\|^2_F=N_t {G}_r,
\end{equation}
$G_r$ is the gain of holographic MIMO array, which is dependent on antenna array topology, beamforming angles, and mutual coupling.
\subsubsection{Tx coherent, Rx coherent}
In this case, correlations at both Tx and Rx are considered, and the channel state information (CSI) is assumed to be known on both sides. For example, in joint Tx-Rx maximum ratio combining (MRC), the receiving power is \cite{larsson2003space}
\begin{equation}
P_r=\sum_{i=1}^{N_t}\left\|\mathbf{h}_i\right\|^2 {P_t}=\|\mathbf{H}\|^2_F {P_t},
\end{equation}
then, according to (\ref{Friis}), we have
\begin{equation}
\|{\mathbf{H}}\|^2_F={G}_t {G}_r,
\end{equation}
where the channel matrix should be normalized according to the antenna array gains at both sides. 
\subsubsection{Tx non-coherent, Rx non-coherent}
The correlations at the Tx and Rx sides are zero, and the CSI is unknown to both sides. The total Rx power can be written as \cite{hochwald2000unitary}
\begin{equation}
P_r=\sum_{i=1}^{N_t}\left\|\mathbf{h}_i\right\|^2 \frac{P_t}{N_t N_r}=\|\mathbf{H}\|^2 \frac{P_t}{N_t N_r}.
\end{equation}
Also combining it with (\ref{Friis}), since $G_t=G_r=1$ due to no correlations between antennas at both sides, the channel matrix should be normalized as (\ref{eq3}), which is the traditional normalization method. This way of normalization is only dependent on antenna number but not other factors like array topology, coupling, etc. 
\subsection{Ergodic model considering antenna effects}
Rather than the quasi-static model, the ergodic capacity of MIMO system in rich multi-path environment is paid more attention when considering practical applications \cite{oestges2006validity}. Resorting to the Kronecker model, the antenna effects are incorporated into the correlation matrices at both sides, such as deformed radiation patterns and reduced antenna efficiencies \cite{xiaoming2013,chen2018review}, which has been well verified through experiments. Assuming the power is equally allocated, the ergodic capacity can be written as
\begin{equation}
\begin{aligned}
C & =\mathbb{E}\left\{\log _2\left[\operatorname{det}\left(\mathbf{I}+\frac{\gamma}{N_t}\mathbf{H}\mathbf{H}^{\dagger}\right)\right]\right\} ,
\end{aligned}
\end{equation}
where $\mathbb{E}$ is the mathematical expectation. The channel matrix is constructed by
\begin{equation}
\mathbf{H}=\left(\mathbf{R}_t^{\frac{1}{2}} \mathbf{H}_w\mathbf{R}_r^{\frac{1}{2}}\right),
\end{equation}
where $\mathbf{R}_t  \in \mathbb{C}_{N_t \times N_t}$ and $\mathbf{R}_r  \in \mathbb{C}_{N_r \times N_r}$ represent the correlation matrices at the Tx and Rx sides, $\mathbf{R}^{\frac{1}{2}}$ represents the Hermitian square root of $\mathbf{R}$, $\mathbf{H}_w  \in \mathbb{C}_{N_t \times N_r}$ denotes the spatially white MIMO channel with i.i.d. complex Gaussian entries. The correlation matrix at the Rx/Tx side can be calculated with
\begin{equation}
\mathbf{R}=\left[\begin{array}{cccc}
1 & c_{12} & \cdots & c_{1 N_{r}} \\
c_{21}^{*} & 1 & \cdots & c_{2 N_{r}} \\
\vdots & \vdots & \ddots & \vdots \\
c_{N_t 1}^{*} & c_{nN_t 2}^{*} & \cdots & 1
\end{array}\right] ,
\end{equation}
where $c_{mn}$ represents the correlation between Tx/Rx antennas $m$ and $n$, and
\begin{equation}
c_{mn}=\frac{\oint \mathcal{Q}_{mn}(\theta,\phi)\sin\theta\mathrm{d} \theta \mathrm{d} \phi}{\sqrt{\oint \mathcal{Q}_{mm}(\theta,\phi) \sin\theta\mathrm{d} \theta \mathrm{d} \phi} \sqrt{\oint \mathcal{Q}_{nn}(\theta,\phi) \sin\theta\mathrm{d} \theta \mathrm{d} \phi}},
\end{equation}
with
\begin{equation}
\begin{aligned}
\mathcal{Q}_{m n}(\theta,\phi)=&\kappa E_{m}^{\theta} (\theta,\phi) E_{ n}^{\theta *}(\theta,\phi)\mathcal{P}^{\theta}(\theta,\phi) +\\&E_{m}^{\phi} (\theta,\phi) E_{ n}^{\phi *}(\theta,\phi)\mathcal{P}^{\phi}(\theta,\phi),
\end{aligned}
\end{equation}
where $\kappa$ is the cross-polar discrimination (XPD), $E^{\theta}(\Omega)$ and $E^{\phi}(\Omega)$ are the $\theta$ and $\phi$ polarized components of antenna radiation pattern, $\mathcal{P}^{\theta}$ and $\mathcal{P}^{\phi}$ denote the angular power spectrum. Therefore, the $\mathbf{H}$ here includes all antenna effects on correlations, and should be normalized in the same way as quasi-static case in each realization, according to the coherence at Tx/Rx side. The $\mathbf{R}_t$ or $\mathbf{R}_r$ is set to $\mathbf{I}$ if the Tx/Rx antennas are uncorrelated. When considering the effect of antenna efficiency, $\mathbf{H}$ should be normalized according to realized gains incorporating efficiency loss due to mismatching and coupling. Alternatively, we can also multiply a loss matrix as $\mathbf{R} \circ \boldsymbol{\Xi}$, $\circ$ is the element-wise product, and
\begin{equation}
\boldsymbol{\Xi}=\sqrt{\mathbf{e}}\sqrt{\mathbf{e}}^{T},
\end{equation}
with entries from embedded radiation efficiencies\cite{Hannan1964}
\begin{equation}
\mathbf{e}=\left[e^{e m b }_1, e^{e m b }_2,  \cdots, e^{e m b }_{n}\right]^{T},
\end{equation}
since the definition of embedded radiation efficiency includes the average of active radiation efficiencies along all radiation directions \cite{Hannan1964}.

The proposed normalization techniques are also applicable for multi-user (MU) MIMO using the Signal-to-Interference-and-Noise-Ratio (SINR) in typical urban scenarios of 3GPP standard \cite{Nicola2022}. In this context, the channel matrix can be constructed using $\mathbf{h}_i$ vectors, each corresponding to a different user. Subsequently, matrix normalization can be performed on the entire channel matrix. Alternatively, vector normalization can be applied individually to each user by setting $\mathbf{h}_i$ to match the gain of the antenna array along the specific direction of the user.

\subsection{Near-field model}
In far-field communications, the normalization of the channel matrix is determined by the far-field gain. For near-field communications, the gain can be similarly defined; however, it is calculated using the dyadic Green's function rather than the far-field scalar Green's function. Generally, the dyadic Green's function can be written as \cite{tsang1985theory}
\begin{equation}\label{17}
\bar{\boldsymbol{{\mathcal{G}}}}\left(\mathbf{r}, \mathbf{r}^{\prime}\right)=\left(\bar{{\mathbf{I}}}+\frac{\nabla \nabla}{k^{2}}\right)g\left(\mathbf{r}, \mathbf{r}^{\prime}\right),
\end{equation}
where $\bar{{\mathbf{I}}}$ is the unit tensor, $k$ is the free-space wavenumber, $g\left(\mathbf{r}, \mathbf{r}^{\prime}\right)$ is the scalar Green's function. The tensor $\bar{\boldsymbol{{\mathcal{G}}}}$ relating the source and field at $\mathbf{r}^{\prime}$ and $\mathbf{r}$ can be rewritten in a matrix form, i.e., 
\begin{equation}
\left[\begin{array}{c}
E_x \\
E_y \\
E_z
\end{array}\right]=\left[\begin{array}{lll}
\mathcal{G}_{x x} & \mathcal{G}_{x y} & \mathcal{G}_{x z} \\
\mathcal{G}_{y x} & \mathcal{G}_{y y} & \mathcal{G}_{y z} \\
\mathcal{G}_{z x} & \mathcal{G}_{z y} & \mathcal{G}_{z z}
\end{array}\right]\left[\begin{array}{c}
J_x \\
J_y \\
J_z
\end{array}\right],
\end{equation}
where three polarizations of source and three polarizations of field are related, thus a $3\times 3$ matrix. Hence, the channel matrix considering full polarizations becomes
\begin{equation}\label{eq19}
\mathbf{H}=\left[\begin{array}{lll}
\mathbf{H}_{x x} & \mathbf{H}_{x y} & \mathbf{H}_{x z} \\
\mathbf{H}_{y x} & \mathbf{H}_{y y} & \mathbf{H}_{y z} \\
\mathbf{H}_{z x} & \mathbf{H}_{z y} & \mathbf{H}_{z z}
\end{array}\right],
\end{equation}
where each sub-matrix $\mathbf{H}_{p q}$ is an $N_t\times N_r$ channel matrix, the subscripts $p$ and $q$ are $x$, $y$, or $z$, denoting the polarizations of field and source. In far-field communications, only $\mathbf{H}_{x x}$ and $\mathbf{H}_{y y}$ are remained, while, strictly, the matrices of all the polarizations should be considered and normalized appropriately in near-field cases. 

For DOF analysis, (\ref{eq19}) is enough without normalization \cite{Shuai2021}. For capacity analysis largely related to channel gain, in order to capture near-field effects, each sub-matrix should be normalized as
\begin{equation}
\|{\mathbf{H}}_{pq}\|^2={G}_t^{pq}{G}_r^{pq},
\end{equation}
where the near-field gain $G^{pq}$ can be calculated following the definition of far-field gain but using the dyadic Green's function. Furthermore, the power density in near field should be calculated with rigorous Poyting vector, since the electric and magnetic fields are no more simply related by a wave impedance. Specific deductions and discussions are put in Section IV. 

In summary, the normalization method of channel matrix should be carefully chosen according to specific scenarios. For holographic MIMO communications, the normalization will definitely dependent on the gain of holographic array, thus introducing extra electromagnetic constraints and insights brought by array topology. Furthermore, the above array gains can be scaled to achieve the desired total power gain while preserving the electromagnetic factors related to the array topology, as discussed in the second paragraph of Section V. In the following sections, we will discuss the gain expressions and limits, utilizing antenna and wave theories. So that we can explore how normalization impacts capacity evaluation when dealing with new array topologies.
\begin{figure}[ht!]
	\centering
	\includegraphics[width=3.4in]{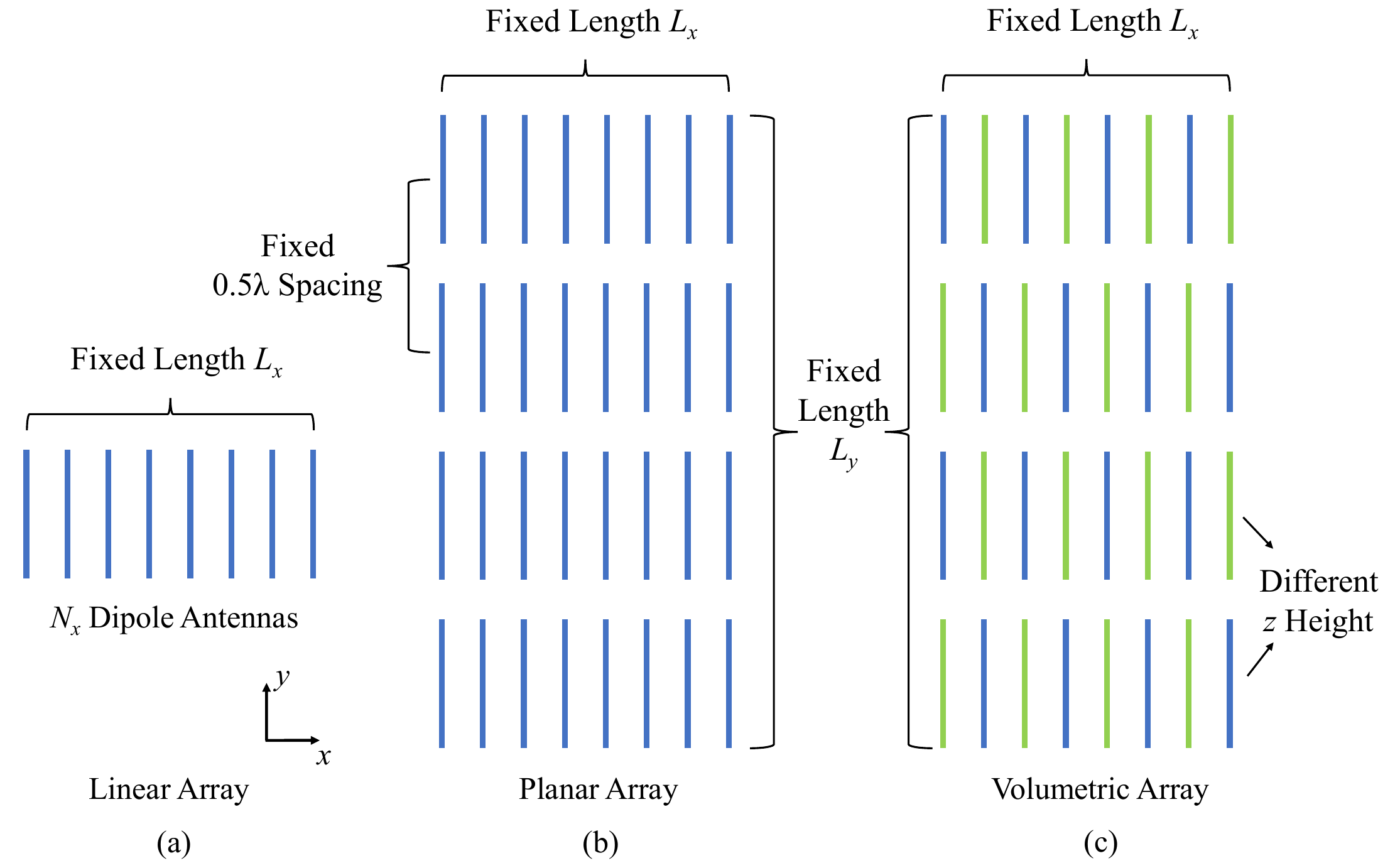}
	\caption{The array topologies considered in this paper. (a) Linear array, the length along $x$ axis is fixed as $L_x$, with a total antenna number $N_x$. (b) Planar array, the length along $x$ and $y$ axes are fixed as $L_x$ and $L_y$, the antenna number along $x$ axis is also $N_x$, and the element spacing along $y$ axis is fixed as 0.5$\lambda$, $\lambda$ is free-space wavelength. (c) Volumetric array, the array topology is the same as planar case, but an extra height difference is introduced along vertical $z$ axis between nearby antennas (denoted by green and blue colors). The $L_x$ and $L_y$ are fixed as 5$\lambda$, and the height difference in volumetric array is fixed as 1$\lambda$ in this paper.}
	\label{gain_linear}
\end{figure}
\section{Far-field gain of arbitrary antenna array}
\subsection{Fundamental analytical formula of gain}
The antenna topologies considered in this paper are depicted in Fig. 1. Generally, the $\theta$ and $\phi$ components of the far-field electric field of the $n$th antenna in an array can be written as $E^{\theta}_n(\theta,\phi)$ and $E^{\phi}_n(\theta,\phi)$, then the total radiation patterns of $N$ antennas are \cite{balanis2015antenna}
\begin{equation}E_{\text{tot}}^{\theta}(\theta,\phi)=\sum_{n=1}^N a_n e^{j[k\Omega_n(\theta,\phi)+\alpha_n]} E_{ n}^{\theta}(\theta,\phi), \end{equation}
\begin{equation}E_{\text{tot}}^{\phi}(\theta,\phi)=\sum_{n=1}^N a_n e^{j[k\Omega_n(\theta,\phi)+\alpha_n]}  E_{ n}^{\phi}(\theta,\phi), \end{equation}
\begin{equation}E_{\text{tot}}= \sqrt{|E_{\text{tot}}^{\theta}|^2+|E_{\text{tot}}^{\phi}|^2} ,\end{equation}
where $k\Omega_n(\theta,\phi) =k{\left(x_n \sin \theta \cos \phi+y_n \sin \theta \sin \phi+z_n \cos \theta\right)} $ is the phase factor of antenna element $n$ at the position $(x_n, y_n, z_n)$ along the direction $(\theta,\phi)$, $a_n$ and $\alpha_n$ are the amplitude and phase of the excitation at antenna $n$, respectively. The far-field radiation intensity is defined as the power radiated per solid angle
\begin{equation}U(\theta, \phi)=\frac{r^2|E_{\text{tot}}|^2}{2\eta},\end{equation}
where $r$ is the far-field reference distance, $\eta$ is wave impedance, and $E=\eta H$ is valid at far field. Then the gain (without ohmic loss) of antenna array can be defined as 
\begin{equation}\label{gain}{G}(\theta, \phi)=\frac{U(\theta, \phi)}{\frac{1}{4 \pi} \int_0^{2 \pi} \int_0^\pi U(\theta, \phi) \sin \theta d \theta d \phi}.\end{equation}
Although the numerator (radiation intensity) in this formula can increase infinitely proportion to antenna number, the denominator introduces electromagnetic constraints from array size, thus the gain limit. When considering the power loss due to antenna mismatching and coupling, the gain becomes realized gain. By performing the integrals over $\theta$ and $\phi$, we can obtain the gain along different directions with designed excitations. However, this process may be complex and time-consuming, especially when a high sampling rate of $\theta$ and $\phi$ is needed, so the closed-form solutions are introduced in next subsection. 
\subsection{Closed-form solution of analytical formula}
\begin{figure}[ht!]
	\centering
	\includegraphics[width=3.4in]{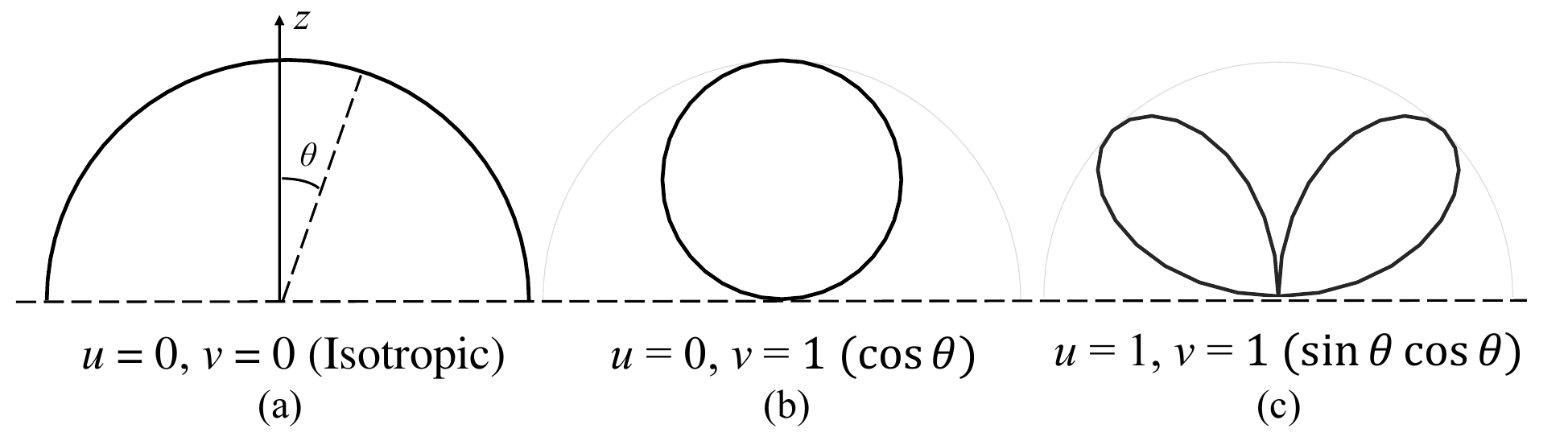}
	\caption{Radiation patterns of array elements with different values of $u$ and $v$. (a) Isotropic pattern of a point source. (b) $\cos$ shape pattern of a dipole antenna with reflection board. (c) Butterfly shape pattern, which is close to the pattern of an orbital-angular-momentum (OAM) antenna.}
	\label{gain_linear}
\end{figure}
General expressions of directivity or gain are useful for antenna synthesis in many scenarios \cite{costa2018closed, das2016generalization,kedar2018wide}. Considering a single polarization, the radiation pattern of an element can be expressed as 
\begin{equation}\label{26}
E(\theta,\phi)=\frac{e^{-jkr}}{r}\sin ^{u} \theta \cos ^{v} \theta,
\end{equation}
with $u>-1, v>-(1 / 2)$, showing rotational symmetry along $\phi$ angle. Three typical patterns are depicted in Fig. 2, i.e., $(u=0, v=0), (u=0, v=1), (u=1, v= 1)$. The radiation patterns of the elements in an array are assumed to be the same, which is valid in most uniform arrays. Take (\ref{26}) into (21-24), the radiation density becomes
\begin{equation}
\begin{aligned}
U= & \frac{1}{2 \eta} \sin ^{2 u} \theta \cos ^{2 v} \theta \sum_{m=1}^N \sum_{n=1}^N a_m a_n e^{j\left(\alpha_m-\alpha_n\right)} \times \\
& e^{j k\left\{\Omega_m(\theta, \phi)-\Omega_n (\theta,\phi) \right\}}\\=& \frac{1}{2 \eta} \sin ^{2 u} \theta \cos ^{2 v} \theta \sum_{m=1}^N \sum_{n=1}^N a_m a_n e^{j\left(\alpha_m-\alpha_n\right)} \times \\
& e^{j k\left\{\rho_{m n} \sin \theta \cos \left(\phi-\phi_{m n}\right)+\left(z_m-z_n\right) \cos \theta\right\}},
\end{aligned}
\end{equation}
where $\rho_{m n}=\sqrt{\left(x_m-x_n\right)^2+\left(y_m-y_n\right)^2}$ and $\phi_{m n}=\tan ^{-1}\left(\frac{y_m-y_n}{x_m-x_n}\right)$. Apparently, the maximum value along one direction can be achieved when all the phases of the excitations are matched, i.e., constructively add along one direction ($\theta_m,\phi_m$) is realized, which becomes
\begin{equation}
U_{\max }=\sin ^{2 u} \theta_m \cos ^{2 v} \theta_m \left(\sum_{n=1}^N a_n\right)^2 / 2 \eta.
\end{equation}
Considering the denominator in (\ref{gain}), the first integral of $U$ over $\phi$ can be deduced as
\begin{equation} \label{P_rad}
\begin{aligned}
& P_{\mathrm{rad}}= \frac{ \pi}{ \eta} \sum_{m=1}^N \sum_{n=1}^N a_m a_n e^{j\left(\alpha_m-\alpha_n\right)}\times \\
&  \int_0^\pi e^{j k\left(z_m-z_n\right) \cos \theta} J_0\left(k \rho_{m n} \sin \theta\right) \sin ^{2 u+1} \theta \cos ^{2 v} \theta d \theta,
\end{aligned}
\end{equation}
which utilizes the property of Bessel function in (\ref{62}) of Appendix A. The exponential term in the integral part of (\ref{P_rad}) can be further expanded as power series
\begin{equation}
\begin{aligned}
e^{j k\left(z_m-z_n\right) \cos \theta}  =  \sum_{p=0}^{\infty} \frac{(j)^p\left(k\left(z_m-z_n\right)\right)^p}{p!},
\end{aligned}
\end{equation}
the rest part including the integral of Bessel function can be obtained using the property of Hypergeometric series in (\ref{63}) of Appendix A as
\begin{equation}
\begin{aligned}
&\int_0^\pi J_0\left(k \rho_{m n} \sin \theta\right) \sin ^{2 u+1} \theta \cos ^{2 v} \theta d \theta= \\& \sum_{p=0}^{\infty}  \frac{\left(1+(-1)^{2 v+p}\right)}{2}  B\left(u+1, v+\frac{p+1}{2}\right) \\
& \times{ }_1 {F}_2\left(\left.\begin{array}{c}
u+1 \\
1, u+v+\frac{p+3}{2}
\end{array} \right\rvert\,-\left(\frac{k \rho_{m n}}{2}\right)^2\right),
\end{aligned}
\end{equation}
where $B$ represents the Beta function of arguments $m$ and $n$
\begin{equation}
\mathrm{B}(m, n)=\frac{(m-1)!(n-1)!}{(m+n-1)!},
\end{equation}
and the hyper-geometric function with order $(1,2)$ is
\begin{equation}
\begin{aligned}
_1 {F}_2\left(\left.\begin{array}{c}
a_1 \\
b_1, b_2
\end{array} \right\rvert\,z\right)=\sum_{k=0}^{\infty} \frac{\left(a_1\right)_k z^k}{\left(b_1\right)_k\left(b_2\right)_k k!},
\end{aligned}
\end{equation}
where $(a)_k=a(a-1)(a-2) \cdots(a-k+1)$ is the falling factorial of $a$, (31) will be valid for any $u>-1,v>-1/2$. Combining (29-33) we can get the closed-form formula for gain with a suitable cut of series, and the results of two typical patterns in Fig. 2 are given below.

\subsubsection{Isotropic pattern $(u=0, v=0)$}
With (\ref{64}) in Appendix A, the radiated power for isotropic antennas can be simplified as
\begin{equation}
P_{\mathrm{rad}}^{\mathrm{iso}}=\frac{2\pi}{ \eta} \sum_{m=1}^N \sum_{n=1}^N a_m a_n e^{j\left(\alpha_m-\alpha_n\right)} \frac{\sin \left(k R_{m n}\right)}{k R_{m n}},
\end{equation}
where $R_{m n}=\sqrt{\left(x_m-x_n\right)^2+\left(y_m-y_n\right)^2+\left(z_m-z_n\right)^2}$.

\subsubsection{$\cos\theta$ shape pattern $(u=0, v=1)$}
Similarly, for $\cos$ shape pattern, the expression of radiated power can be simplified as
\begin{equation}
\begin{aligned}
&P_{\mathrm{rad}}^{\mathrm{cos}}=\frac{2\pi}{\eta}\sum_{n=1}^N \frac{a_n^2}{3}-\frac{4\pi}{\eta} \sum_{\substack{n, m=1 \\ m \neq n \\ n>m}}^N a_n a_m e^{j\left(\alpha_m-\alpha_n\right)}\times \\&\left[\frac{\left(\rho_{mn}^2-2 z_{m n}^2\right) \cos \left(kR_{mn}\right)}{k^2R_{mn}^4}-\right.\\&\left.\frac{\left(\left(k^2\rho_{mn}^2-2\right) z_{m n}^2+\rho_{mn}^2+k^2z_{m n}^4\right) \sin \left(kR_{mn}\right)}{k^3R_{mn}^{5}}\right].
\end{aligned}
\end{equation}
\subsection{Physical formula of gain}
The physical limit of antenna array gain is dependent on the effective aperture of array \cite{stegen1964gain, nuttall2001approximations, gustafsson2024shadow}, and the formula is
\begin{equation}
{G}=4\pi A_e/\lambda^2 ,
\end{equation}
where effective area $A_e$ is related to array topology and beamforming direction. Particularly, for MIMO applications, rather than the gain along a specific direction, we need to consider the average gain inside the target angular spread $\theta = 0 \sim \theta_0, \phi = 0\sim \phi_0$. Therefore, we can use the average effective area
\begin{equation}
\hat{A_e}=\frac{\int_{0}^{\phi_0}\int_{0}^{\theta_0}A(\theta,\phi) d\theta d\phi}{\theta_0\phi_0},
\end{equation}
where $\hat{A_e}$ here denotes the average effective area. $\theta_0 = 60^{\circ}$ and $\phi_0= 0 ^{\circ}$ or $180^{\circ}$ are used in this paper, which indicates the most representative $\pm 60^{\circ}$ horizontal scanning in urban base-station applications. Average gains within different scanning ranges can also be easily obtained with this method. The effective apertures of linear, planar, and volumetric arrays are discussed below.
\subsubsection{Linear array}
Traditionally, the gain limit of linear array is $2L/\lambda$, and the concept of `area' is not introduced in this formula or any analytical method. However, in practical scenarios, a linear array does have a certain effective area, which can lead to discrepancies between this theoretical formula and full-wave simulations. Specifically, both two dimensions of practical antenna array will contribute to gain, and the decrease of gain along $\theta$ cannot be captured without the concept of `area'. In our approach, we address this issue by treating the linear array as a planar array with an effective length of  $0.68\lambda$ along the $y$ axis. This adjustment helps better align theoretical predictions with full-wave simulation results by incorporating the effective area of the array. Hence, the effective area of a linear array becomes
\begin{equation}\label{38}
A_e= 0.68\lambda\times L \cos\theta,
\end{equation}
which makes the gain results match with full-wave simulations. The average gain in a certain $\theta$ and $\phi$ range is
\begin{equation}
\hat{A_e}= \frac{ 0.68\lambda\times L\sin\theta_0}{\theta_0}.
\end{equation}
\subsubsection{Planar array}
For planar array, we have the effecitve area limit
\begin{equation}
A_e=L_x L_y\cos\theta,
\end{equation}
and the average gain is
\begin{equation}
\hat{A_e}= \frac{ L_x L_y\sin\theta_0}{\theta_0}.
\end{equation}
\subsubsection{Volumetric array}
For volumetric array, the effective area is the add of the projection or shadow areas along three dimensions, which is
\begin{equation}
\begin{aligned}
A_e =& L_x L_y \cos \theta +L_x L_z \sin \theta \sin \phi +\\&L_y L_z \sin \theta\cos \phi,
\end{aligned}
\end{equation}
where $L_z$ is the vertical length of the volumetric array, and the average gain is
\begin{equation}
\begin{aligned}
\hat{A_e}= &\frac{ L_x L_y\sin\theta_0}{\theta_0}+\frac{ L_x L_z(1-\cos\theta_0) (1-\cos\phi_0) }{\theta_0\phi_0}\\&+\frac{ L_y L_z(1-\cos\theta_0) \sin\phi_0 }{\theta_0\phi_0}.
\end{aligned}
\end{equation}
\subsection{Physical formula of realized gain}
In practical antenna array analysis, the realized gain, accounting for antenna efficiency losses due to mismatching and coupling, cannot be neglected \cite{Hannan1964, ji2024}. For dense array, the embedded radiation efficiency of an array element can be approximately estimated with \cite{Hannan1964}
\begin{equation}
e=\frac{G_e}{D_e}=\frac{4 \pi S_e}{D_e\lambda^2},
\end{equation}
where $G_e$, $D_e$, and $S_e$ are the gain, directivity, and area of a single element, respectively. For an array element, the gain decreases with a reduction in its area, while its directivity remains unchanged for small elements, such as a short dipole. Consequently, any decrease in gain must be attributed to efficiency loss, which leads to the formula. This formula is quite accurate for planar arrays, as demonstrated by previous studies \cite{ShuaiOJAP, Kildal2016}, but it needs modifications for linear and volumetric arrays. In this work, the directivity of the array element is set to $D_e=3.28$, which doubles that of a standard dipole. The efficiency expressions used in this study are provided below as
\begin{equation} 
e = \begin{cases}
a_l\sqrt{4\pi S_e/(D_e\lambda^2)} &\text{Linear},\\
4 \pi S_e/(D_e\lambda^2) &\text{Planar},\\ 
4 \pi (S_e+S_v)/(D_e\lambda^2)    & \text{Volumetric},
\end{cases} [e\le 1].
\end{equation}
For the planar array considered here, the $y$ dimension of each element is fixed at $0.5\lambda$, making the area of a single element $S_e =0.5\lambda\times L_x/N_x$, $N_x$ is the antenna number along $x$ axis. In the case of a volumetric array, each element can be regarded to have an equivalent larger area due to the additional vertical dimension, denoted as $S_v$. Since the antenna spacing along the vertical dimension is not affected by $N_x$, $S_v$ is treated as a fixed value, and retrieved as $0.065\lambda^2$, nearly half of the element area at $0.25\lambda$ spacing, to align well with full-wave simulations. For linear arrays, the coupling effect is less pronounced because only antennas along one dimension contribute to mutual coupling, resulting in a more moderate decrease in efficiency. Here, efficiency is represented as the square root of that for a planar array, with a retrieved coefficient $a_l = 0.77$ to account for specific antenna designs. The element area for the element in linear array is $S_e =0.68\lambda\times L_x/N_x$, as discussed in (\ref{38}). It is important to note that the efficiency models presented here are influenced by antenna design, excitation strategy, and array topology, which may result in different values of $a_l$, $S_e$, $S_v$, and $D_e$. Nonetheless, the planar cases generally provide accurate results, and the formulae here for both linear and volumetric arrays show good agreement with full-wave simulations using standard dipoles.
\begin{figure}[ht!]
	\centering
	\includegraphics[width=3in]{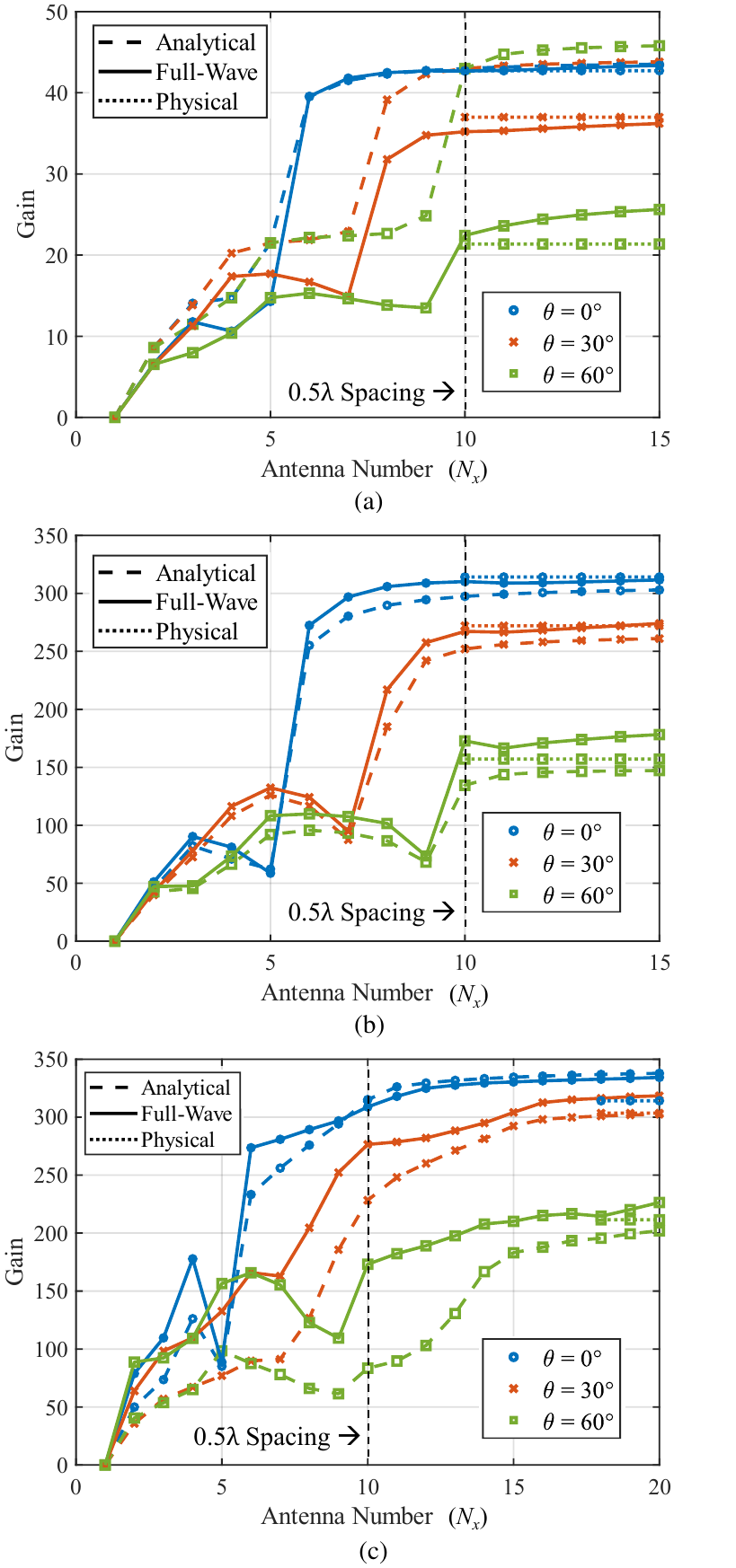}
	\caption{Gains of antenna arrays calculated with the analytical, physical, and full-wave methods at beamforming angles $\theta = 0^{\circ}, 30^{\circ}, 60^{\circ}$. The antenna number at $0.5\lambda$ element spacing along the $x$ axis is denoted by the black dotted line. (a) Linear array. (b) Planar array. (c) Volumetric array.}
	\label{gain_linear}
\end{figure}
\subsection{Numerical verifications}
\begin{figure}[ht!]
	\centering
	\includegraphics[width=3in]{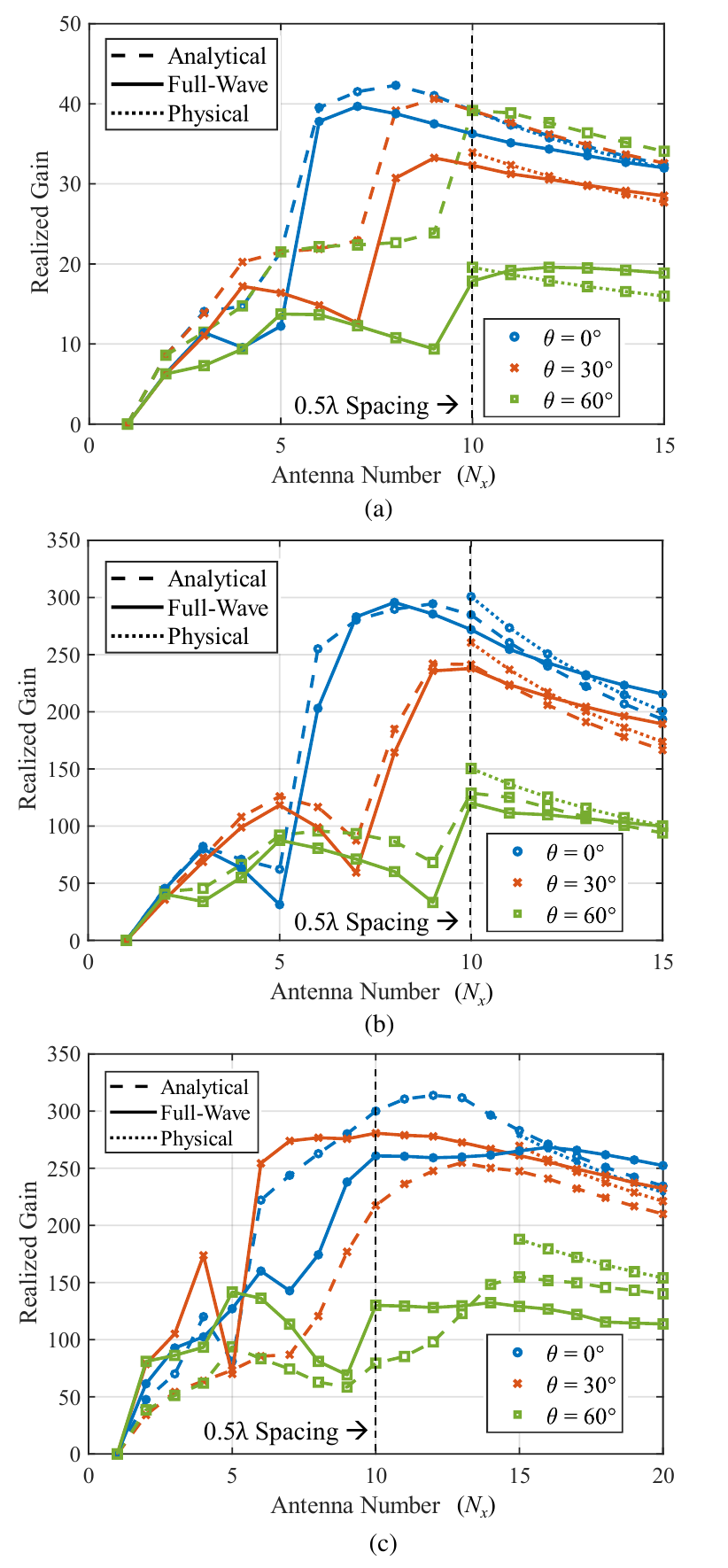}
	\caption{Realized gains of antenna arrays calculated with analytical, physical, and full-wave methods at beamforming angles $\theta = 0^{\circ}, 30^{\circ}, 60^{\circ}$. The antenna number at $0.5\lambda$ element spacing along $x$ axis is denoted by the black dotted line. (a) Linear array. (b) Planar array. (c) Volumetric array.}
	\label{gain_volumetric}
\end{figure}
\subsubsection{Configuration}
The linear, planar, and volumetric array topologies are depicted in Fig. 1. To investigate the gain properties of different array topologies, we compare the results from analytical methods, physical methods, and full-wave simulations as follows:
\begin{itemize} 
	\item Analytical Method: Closed-form formulae (34-35) can be used for obtaining the analytical results, and $+-60^\circ$ horizontal scanning is considered here with standard dipole antenna array.
	\item Physical Method: Formulae (38-43) are used to compute the gain limits, and formula (45) is utilized to determine the antenna efficiencies, thus realized gain results, based on the gain calculations.
	\item Full-Wave Method: Full-wave simulations are conducted using CST Microwave Studio (CST MWS), employing a basic micro-strip dipole antenna as the array element \cite{ShuaiOJAP}. 
\end{itemize} 

Since the analytical and full-wave cases consider antennas without reflective boards, the gain results from these two methods are multiplied by 2 to align with the physical results \cite{balanis2015antenna}, which assume the presence of an ideal reflective board. To ensure fair comparisons, the beamforming patterns generated are made strictly symmetric across the $xoy$ plane for all antenna arrays in the analytical and full-wave cases.
\begin{figure}[ht!]
	\centering
	\includegraphics[width=3in]{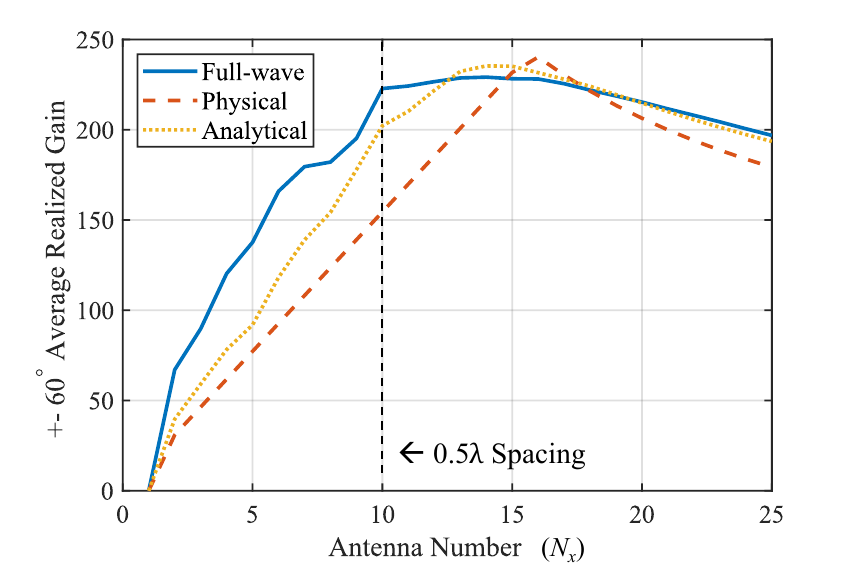}
	\caption{Average realized gains of volumetric array within angular spread $\theta = 0 \sim 60^{\circ}$, i.e., $\pm 60^{\circ}$ horizontal scanning, calculated with analytical, physical, and full-wave methods.}
	\label{gain_volumetric}
\end{figure}
\subsubsection{Gain results}
The gain results are illustrated in Fig. 3. In the linear case, the analytical method fails to capture the decrease in gain at larger radiation angles because it does not account for the concept of `area' in its formulae—an often overlooked limitation in many analyses. In contrast, the proposed physical method aligns well with full-wave simulations, underscoring the importance of considering the antenna's effective area, even in linear cases. In the planar case, all three methods closely agree, accurately characterizing the gain limits across different beamforming directions. For the volumetric array, the analytical and full-wave methods show good consistency, especially when the number of antennas is sufficient to properly sample the array volume. Gains calculated using the physical method are slightly lower compared to the other two methods. Although the effective aperture of the volumetric array is considered the same as that of the planar array along the vertical direction, small benefit relating to vertical length can still be observed with the analytical or full-wave method. Notably, the gain of the volumetric array significantly surpasses that of the planar array with the same aperture size at larger angles, which aligns with physical analysis and highlights the advantages of exploring volumetric array topologies. Additionally, the optimal number of antennas varies with the beamforming direction, with more antennas required for larger beamforming angles.
\subsubsection{Realized gain results}
The results for realized gains, incorporating antenna efficiency effects, are depicted in Fig. 4. Across all cases, the gain decreases with smaller element spacing due to reduced efficiency. This decrease is more gradual for linear and volumetric arrays compared to planar arrays, as the effective area of a single element diminishes more rapidly in planar arrays. For linear and planar cases, the three models generally agree well, although the analytical results for the linear case show discrepancies similar to those observed in the gain discussions. Estimating the realized gain for volumetric arrays at each beamforming angle proves challenging with physical or analytical methods due to the complexity of active impedance. However, when considering the average realized gain within a target angular spread, the gains calculated using physical and analytical methods align well with full-wave simulations, particularly with the closed-form method, as shown in Fig. 5. Thus, the proposed efficiency formulae are effective for linear and planar cases and also yield satisfactory results for volumetric arrays when considering average gain.
\section{Near-field gain of arbitrary antenna array}
The gains discussed previously are applicable to the normalization of the channel matrix in far-field communications. However, near-field communications are becoming increasingly important in current 6G frameworks \cite{lu2021communicating, wang2024tutorial}. Building on the far-field gain definition, we derive the formula for near-field gain and reveal that additional gain losses may arise in near-field communications.
\subsection{Rigorous analysis}
From the perspective of electromagnetic theory, the differences between near-field and far-field communications primarily involve the following three aspects: (a) Distance effects must be considered, as field strength does not decrease with $1/r$ in near field; (b) Cross-polarizations can be generated from a single polarization excitation at close distances; and (c) When calculating power intensity, the strict Poynting vector must be used, as the relationship $E=\eta H$ does not hold in the near field. 

The scalar Green's function is
\begin{equation}\label{46}
g\left(\mathbf{r}, \mathbf{r}^{\prime}\right)=\frac{1}{4 \pi} \frac{\exp \left(-j k\left|\mathbf{r}-\mathbf{r}^{\prime}\right|\right)}{\left|\mathbf{r}-\mathbf{r}^{\prime}\right|},
\end{equation}
and $R=\left|\mathbf{r}-\mathbf{r}^{\prime}\right|$. At far-field, $R \approx r-\mathbf{r}^{\prime} \cdot \mathbf{a}_{\mathrm{R}}$, $\mathbf{a}_{\mathrm{R}}$ is the unit vector along $\mathbf{r}-\mathbf{r}^{\prime}$. Hence, the scalar Geen's function under far-field approximation becomes
\begin{equation}\label{47}
g_{\text {far }}\left(\mathbf{k}_R, \mathbf{r}^{\prime}\right)=\frac{e^{-j k r}}{4 \pi r} \exp \left(j k \mathbf{r}^{\prime} \cdot \mathbf{k}_R\right),
\end{equation}
which is dependent on the direction vector $\mathbf{k}_R$ but not position $\mathbf{r}$. For near-field analysis, we have electric and magnetic dyadic Green's function 
\begin{equation}
\boldsymbol{\mathcal{E}}=-j \omega \mu_0 \int_V  \bar{\boldsymbol{\mathcal{G}}}\left(\mathbf{r}, \mathbf{r}^{\prime}\right) \cdot \boldsymbol{\mathcal{J}}\left(\mathbf{r}^{\prime}\right) d \mathbf{r}^{\prime},
\end{equation}
\begin{equation}
\boldsymbol{\mathcal{H}}=\nabla \times \int_V \bar{\boldsymbol{\mathcal{G}}}\left(\mathbf{r}, \mathbf{r}^{\prime}\right) \cdot \boldsymbol{\mathcal{J}}\left(\mathbf{r}^{\prime}\right) d \mathbf{r}^{\prime},
\end{equation}
where $\omega$ and $\mu$ are the angular frequency and free-space permeability, $\boldsymbol{\mathcal{J}}$ denotes equivalent current. From (\ref{17}), we have electric Green's function
\begin{equation}
\bar{\boldsymbol{\mathcal{G}}}\left(\mathbf{r}, \mathbf{r}^{\prime}\right)=\left(\begin{array}{ccc}
k^2+\frac{\partial^2}{\partial x^2} & \frac{\partial^2}{\partial x \partial y} & \frac{\partial^2}{\partial x \partial z} \\
\frac{\partial^2}{\partial y \partial x} & k^2+\frac{\partial^2}{\partial y^2} & \frac{\partial^2}{\partial y \partial z} \\
\frac{\partial^2}{\partial z \partial x} & \frac{\partial^2}{\partial z \partial y} & k^2+\frac{\partial^2}{\partial z^2}
\end{array}\right) \frac{e^{-j kR}}{4 \pi k^2R},
\end{equation}
and magnetic Green's function
\begin{equation}
\begin{aligned}
\nabla \times  \bar{\boldsymbol{\mathcal{G}}}\left(\mathbf{r}, \mathbf{r}^{\prime}\right) &=\nabla \times\left[\left(\bar{{\mathbf{I}}}+\frac{1}{k^2} \nabla \nabla\right) g\left(\mathbf{r}, \mathbf{r}^{\prime}\right)\right] \\&=\left(\begin{array}{ccc}
0 & -\frac{\partial}{\partial z} & \frac{\partial}{\partial y} \\
\frac{\partial}{\partial z} & 0 & -\frac{\partial}{\partial x} \\
-\frac{\partial}{\partial y} & \frac{\partial}{\partial x} & 0
\end{array}\right) \frac{e^{-j k R}}{4 \pi R}
\end{aligned}.
\end{equation}
More explicit forms of the Green's functions can be found in Appendix B for easily calculating the results. 

For $p$ polarization of source ($p, q = x, y, z$), it will generate vector electric and magnetic fields $\boldsymbol{\mathcal{E}}_p$ and $\boldsymbol{\mathcal{H}}_p$, then the Poyting vector can be written as
\begin{equation}
\mathbf{S}_p=\frac{1}{2} \operatorname{Re}\left[\boldsymbol{\mathcal{E}}_p(\mathbf{r}) \times \boldsymbol{\mathcal{H}}_p^*(\mathbf{r})\right],
\end{equation}
representing the power flows as
\begin{equation}
\begin{aligned}
\boldsymbol{\mathcal{E}}_p\times \boldsymbol{\mathcal{H}}_p^* =& \left(\mathcal{E}_y H_z^*-\mathcal{E}_z \mathcal{H}_y^*\right) \hat{\mathbf{x}}+\left(\mathcal{E}_z H_x^*-\mathcal{E}_x \mathcal{H}_z^*\right) \hat{\mathbf{y}}+\\ &\left(\mathcal{E}_x \mathcal{H}_y^*-\mathcal{E}_y \mathcal{H}_x^*\right) \hat{\mathbf{z}} \\=& \left(\mathcal{E}_y \mathcal{H}_z^*-\mathcal{E}_z \mathcal{H}_y^*\right) \mathbf{y}+\left(\mathcal{E}_z \mathcal{H}_x^*-\mathcal{E}_x \mathcal{H}_z^*\right) \mathbf{z}+\\ &\left(\mathcal{E}_x \mathcal{H}_y^*-\mathcal{E}_y \mathcal{H}_x^*\right) \mathbf{x},
\end{aligned}
\end{equation}
where $\hat{\mathbf{x}}, \hat{\mathbf{y}}, \hat{\mathbf{z}}$ are the direction of power flow, and we re-write it in consistent of the field polarization ${\mathbf{x}}, {\mathbf{y}}, {\mathbf{z}}$, i.e., the ${\mathbf{x}}$ component is dominated by generated $x$-polarized field. We can further denote the $q$ component of $\mathbf{S}_p$ as $S_{pq}$. The total radiated power is then obtained with the integral of Poyting vector over surface area $S_a$
\begin{equation}
P_{\text{total}}=\int_{S_a} \frac{1}{2} \operatorname{Re}\left[\boldsymbol{\mathcal{E}}(\mathbf{r}) \times \boldsymbol{\mathcal{H}}^*(\mathbf{r})\right] \cdot d \mathbf{S}_a.
\end{equation}
According to power conservation, total radiated power should be the same at near and far-field, so we can also use the far-field radiated power as $P_{\text{total}}$.

Now, we can define near-field gain following similar route of far-field gain. Assuming there are $N$ antennas focusing at position $\mathbf{r}_f$, for the excitation of polarization $p$ with amplitude $a_n$ and phase $\alpha_n$, the generated $q$-polarized electric field can be calculated with
\begin{equation}\label{55}
\mathcal{E}_{q}(\mathbf{r}_f)=\sum_{n=1}^{N} a_n e^{j\alpha_n} \mathcal{G}_{qp}(\mathbf{r}_f,\mathbf{r}^{\prime}_n),
\end{equation}
and the magnetic field can be calculated in the same way by using the magnetic Green's function. Then, the Poyting vector from an antenna array can be calculated following (52-53). We can still define the radiation intensity as the power per solid angle at near field, although it is usually a far-field concept, which gives
\begin{equation}
U_{pq}(\theta, \phi,R)= r^2 S_{p q} (\theta, \phi,R),
\end{equation}
where $r$ is the reference distance between source and receiving planes, $S_{pq}$ denotes the $q$ component of Poyting vector generated from source of $p$ polarization. Similar to far-field gain, the near-field gain corresponding to different polarizations can be defined as
\begin{equation}\label{57}G^{pq}=\frac{4\pi U_{pq}(\theta, \phi,R)}{P_{\text{total}}},\end{equation}
the denominator is the total radiated power. Then the channel matrix $\mathbf{H}_{pq}$ can be normalized according to the gain ${G}^{pq}$ considering near-field effects, as discussed in Section II.
\subsection{Losses in near-field gain}
Near-field gain can experience additional losses beyond the fundamental far-field gain limits imposed by effective area and antenna efficiency. To explore these characteristics, we calculate near-field gains using both dyadic and scalar Green's functions, applying either near-field or far-field beamforming. Specifically, (\ref{57}) deduced from dyadic Green's function is used as the dyadic approach. For the scalar approach, the $\mathcal{G}_{qp}$ in (\ref{55}) is replaced by the scalar Green's function (\ref{46}) for calculating the near-field radiation intensity and gain. For near-field beamforming, the fields generated by the Tx antennas are set in phase at the target position through appropriate excitations. In contrast, far-field beamforming aligns the fields in phase along the direction of Rx, rather than focusing on the specific Rx position. Both Tx and Rx polarizations are set to $x$ in the numerical examples considered here.

\begin{figure}[ht!]
	\centering
	\includegraphics[width=3.4in]{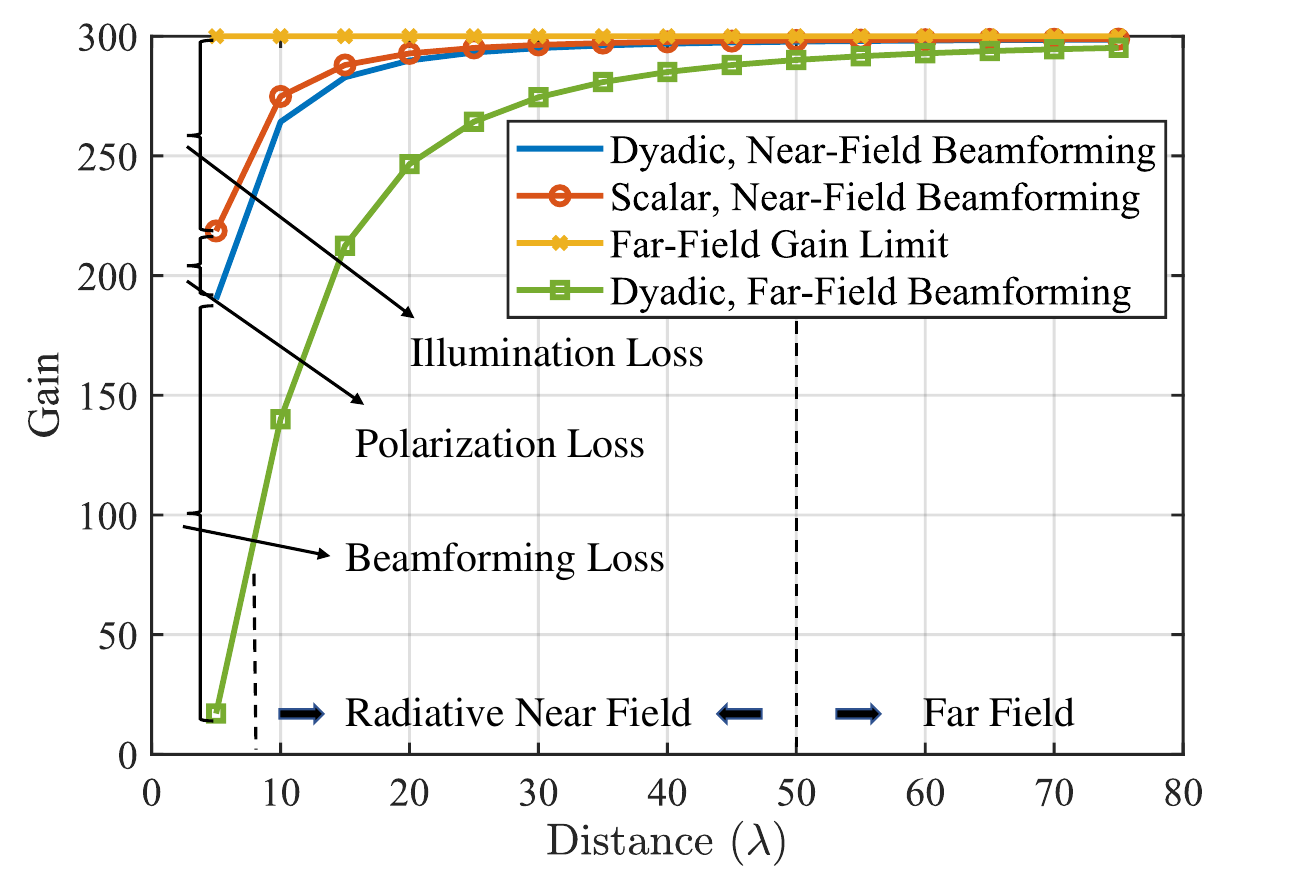}
	\caption{Near-field gains of antenna arrays calculated using dyadic and scalar Green's functions with near-field or far-field beamforming. The differences between gains at each distance can be attributed to the different loss factors considered in these calculation methods, where all the gains will tend to far-field gain limit at large distance.}
	\label{gain_volumetric}
\end{figure}
\begin{figure}[ht!]
	\centering
	\includegraphics[width=3.4in]{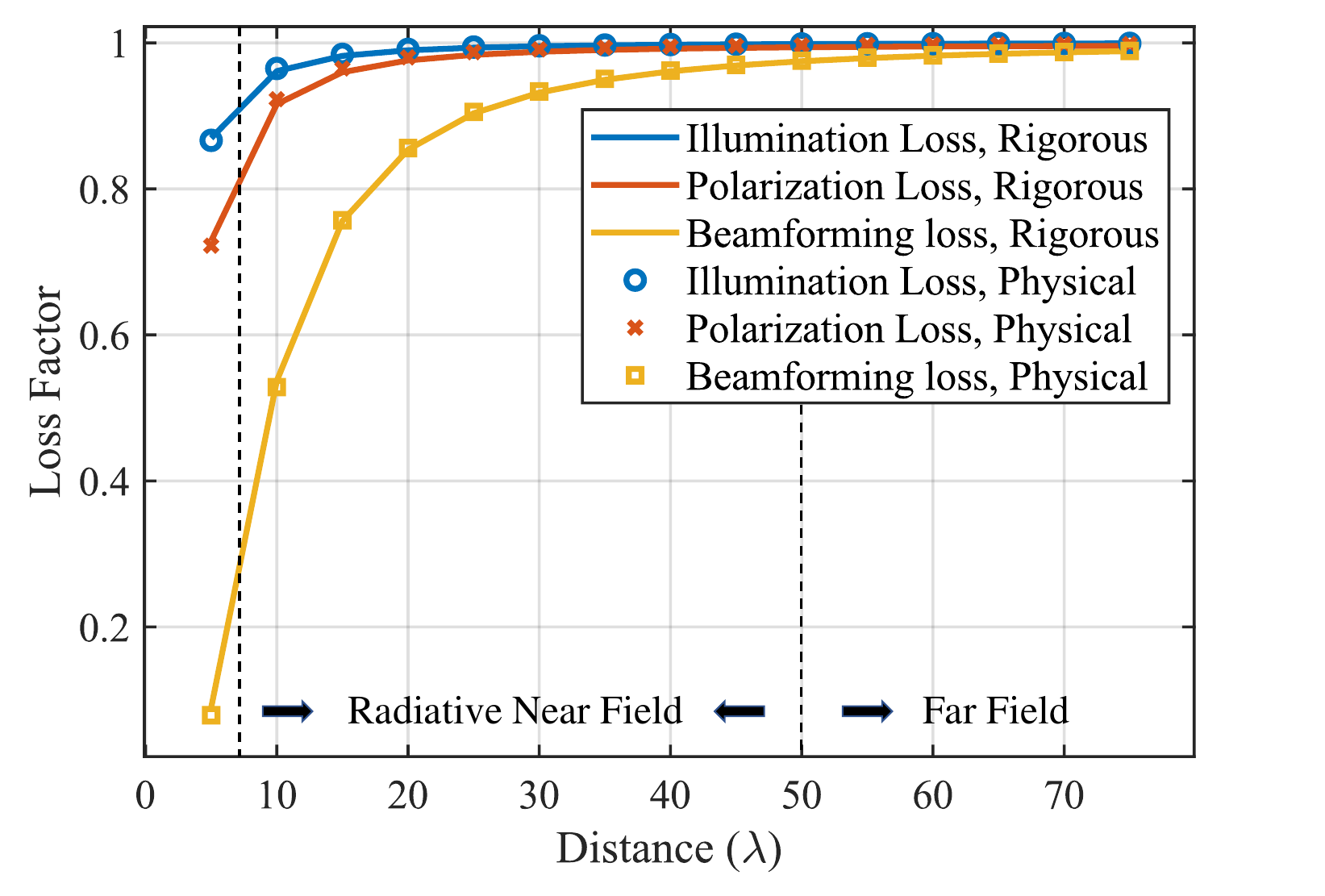}
	\caption{Three loss factors of near-field gains at different distances calculated with rigorous and physical methods.}
	\label{gain_volumetric}
\end{figure}
As shown in Fig. 6, the differences in gains obtained from these methods are due to additional losses specific to the near-field environment, including polarization loss, illumination loss, and beamforming loss. Polarization loss occurs when power is converted into undesired polarizations. Illumination loss arises from non-uniform field distribution when the array is illuminated by a source in the near field, and vice versa. This type of loss can be mitigated through amplitude control or even excluded from the near-field gain definition. Beamforming loss occurs when element excitations are optimized for directional far-field beamforming rather than specific near-field positions, a common issue in aperture antenna analysis \cite{kay1960near}. Polarization loss in the near field is almost unavoidable and currently can only be effectively suppressed by elimination techniques, such as oblique projection filter and zero-forcing method, which requires additional computational cost and the manipulation of $z$ polarization \cite{wei2023tri}. Illumination and beamforming losses can be managed through appropriate excitations. To account for these additional losses easily, a physical loss factor can be incorporated into the far-field gain.
\subsection{Physical loss factor}
The previous deductions are quite complex. In practice, simpler physical methods can be used, particularly those based on the near-field gain of aperture antennas. In the heuristic derivation of the near-field range equation, an efficiency factor is introduced to account for gain loss in aperture antennas at near-field distances \cite{kay1960near}. Inspired by this approach, we propose a physical loss factor to address the various losses discussed. For aperture antennas with aperture field distribution
\begin{equation}
f(r)=e^{-j k r / 2}\left[\left(1-\left(r_1 / r_0\right)^2\right)^m+c_m\right],
\end{equation}
where $r_1$ is the distance between field point and central point on the aperture, $r_0$ is the radius of the aperture, $r$ is the reference distance between source and receiver, $m, c_m$ are the factors determining the field distribution of the aperture antenna. From the definition of power gain at arbitrary distance in aperture antenna theory, the loss factor of near-field gain can be approximately calculated with \cite{kay1960near, xiao2021near}
\begin{equation}
\eta=\frac{\sigma^2\left|I(m, m)+c_m I(m, 0)+c_m I(0, m)+c_m^2 I(0,0)\right|^2}{16\left(\frac{c_m{ }^2}{2}+\frac{c_m}{m+1}+\frac{1}{4 m+2}\right)^2},
\end{equation}
where $\sigma$ is a factor related to aperture size and reference distance, $I(m,n)=n!m!\sum_{j=0}^{\infty}\left[(-1)^j(\sigma / 2)^{2 j}\right] /[(j+m+1)!(j+n+1)!]$. With uniform excitations ($c_m=0, m=0$), the loss factor is simplified to
\begin{equation}
\eta=\left|\sum_{j=0}^{\infty} \frac{(-1)^j(\sigma)^{2 j}}{(j+1)!(j+1)!}\right|.
\end{equation}
For square arrays, the expression for $\sigma$ can be approximated as $a_L kL/R$, where $L$ is the side length of array, and $a_L$ is the coefficient accounting for the loss considered. The $a_L$ can be retrieved from rigorous analysis as follows
\begin{equation} 
a_L = \begin{cases}
0.12 &\text{Polarization loss},\\
0.18 &\text{Illumination loss},\\
0.5   & \text{Beamforming loss}.
\end{cases}
\end{equation}
The verifications of these physical loss factors are provided in Fig. 7, where great matches can be observed. This approach facilitates the straightforward adaptation of the far-field gain to estimate the near-field gain by incorporating the appropriate loss factors.

Therefore, simple equations can be used to account for additional gain losses in near-field communications, if necessary, as these losses may impact capacity evaluations in specific scenarios. In many cases, using far-field gain or employing strict methods without normalization remains appropriate for near-field communications. It is also important to note that the angular spread differs in near-field communications, which suggests that the average gain could differ from that in typical far-field cases.
\section{Effects of normalization on capacity estimation}
\begin{figure}[ht!]
	\centering
	\includegraphics[width=3in]{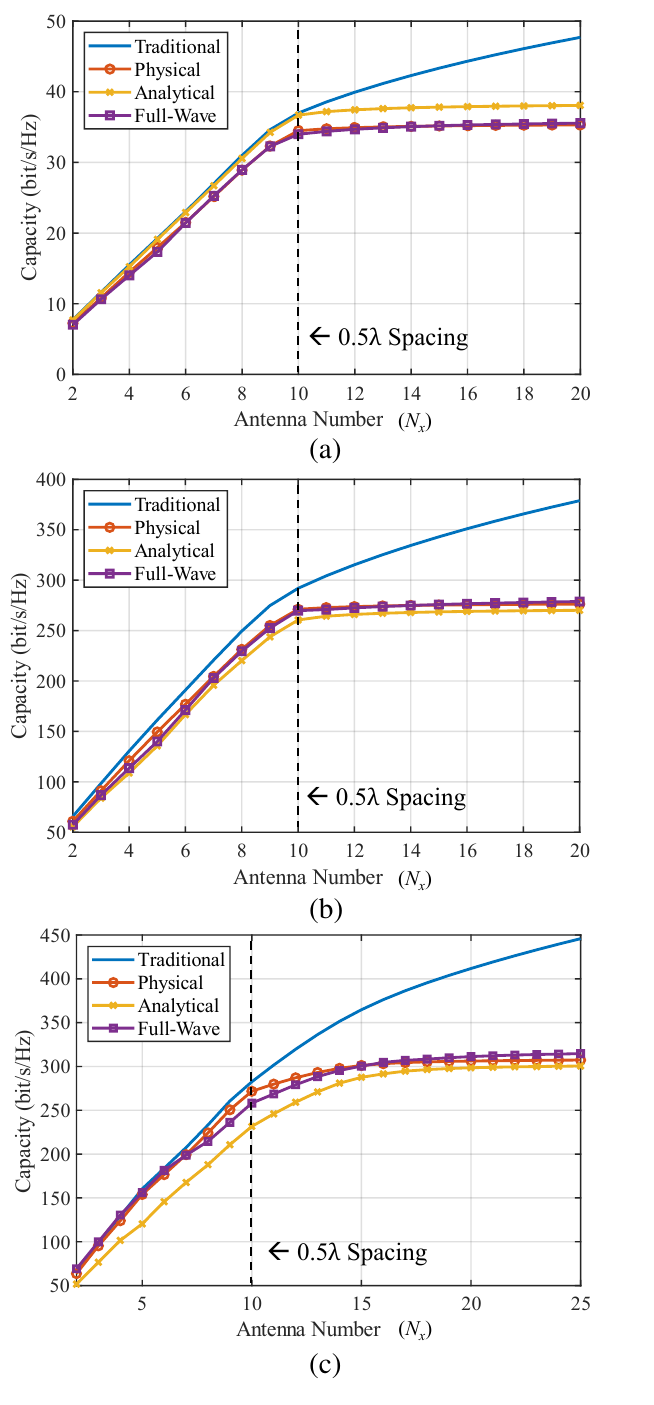}
	\caption{Far-field holographic MIMO capacities using different channel normalization methods in quasi-static cases. (a) Linear array. (b) Planar array. (c) Volumetric array.}
	\label{gain_volumetric}
\end{figure}
The choice of normalization method significantly impacts the performance evaluation of novel antenna array topologies, such as dense and volumetric arrays. In this section, we compare the capacities of linear, planar, and volumetric arrays using the proposed normalization methods across quasi-static, ergodic, and near-field scenarios. For the quasi-static and ergodic scenarios, we employ the BLAST strategy: the antennas at the base station (holographic array) are correlated, while the antennas on the user side remain uncorrelated. The holographic array configurations illustrated in Fig. 1 are used, with the number of users set to 10 for the linear array and 100 for the planar or volumetric array. To highlight the influence of antenna configurations, antenna efficiencies are excluded from the quasi-static cases but are included in the ergodic cases. Correlations between antennas are derived using (12-14) in a rich multi-path environment, with an angular spread of $\pm 60^{\circ}$ aligning with the desired beamforming directions, where uniform power spectrum is assumed and the XPD is 1. For the near-field scenario, a holographic planar or volumetric array of size $5\lambda \times 5\lambda$ is used as the Rx, with the number of antennas reaching its capacity limit, while a standard planar Tx array of the same size and 0.5$\lambda$ element spacing is positioned directly opposite at a distance of $5\lambda$.

Additionally, the antenna array gains for all cases are scaled down by dividing $\pi$, which makes the electromagnetic normalization (\ref{6}) of a planar array close to $||\mathbf{H}||_F = N_t \frac{4A_e}{\lambda^2} = N_t N_r$ at $0.5\lambda$ spacing. With this scaling, the electromagnetic normalization of a planar array approximates a unit sub-channel gain, making comparisons with traditional methods more straightforward. At the same time, the trends influenced by electromagnetic factors are preserved, including the gain limits of the arrays, efficiency reductions due to coupling, and the advantages offered by volumetric arrays. The scaling of the SNR can be adjusted according to the desired total power gain level, while the proposed methods ensure that the relevant electromagnetic factors are accurately captured.
\subsection{Quasi-static capacity}
The capacities in the quasi-static scenario are illustrated in Fig. 8. In all cases using traditional normalization, capacity increases with the number of antennas, which is a non-physical phenomenon resulting from the neglect of the gain limit. For the linear array, the analytical method predicts higher capacities than the physical and full-wave methods, as it fails to account for the gain reduction at increasing $\theta$ angles. In the planar and volumetric arrays, the three proposed normalization methods show good agreement, demonstrating that each method is suitable for normalization. The advantages of volumetric arrays are also evident, as introducing height variations improves gain at large beamforming angles. The planar array approaches its capacity limit at a 0.5$\lambda$ spacing, whereas the capacity of the volumetric array with the same aperture size continues to increase until reaching a 0.25$\lambda$ spacing, ultimately achieving a $15\%$ capacity increase compared to planar array. Notably, without electromagnetic normalization, the correct trend of capacity cannot be accurately captured, and only a $9\%$ benefit will be observed for the volumetric array.

\subsection{Ergodic capacity considering antenna effects}
For the ergodic results shown in Fig. 9, we further account for the impact of antenna efficiency due to mutual coupling. When using traditional normalization methods, the results suggest unlimited capacity gains as the number of antennas increases, even when antenna efficiency is reduced. However, with the proposed normalization method, the capacities of linear and planar arrays reach their limit at 0.5$\lambda$ spacing and then start to decline. This decline is attributed to reduced antenna efficiency, which results in a decrease in realized gain. In contrast, the volumetric array, benefiting from additional DOF, reaches its capacity limit at approximately 0.3$\lambda$ spacing and achieves a $20\%$ capacity gain compared to the planar array. Nevertheless, under traditional normalization, a capacity gain of only $ 13 \%$ is observed.
\begin{figure}[ht!]
	\centering
	\includegraphics[width=3in]{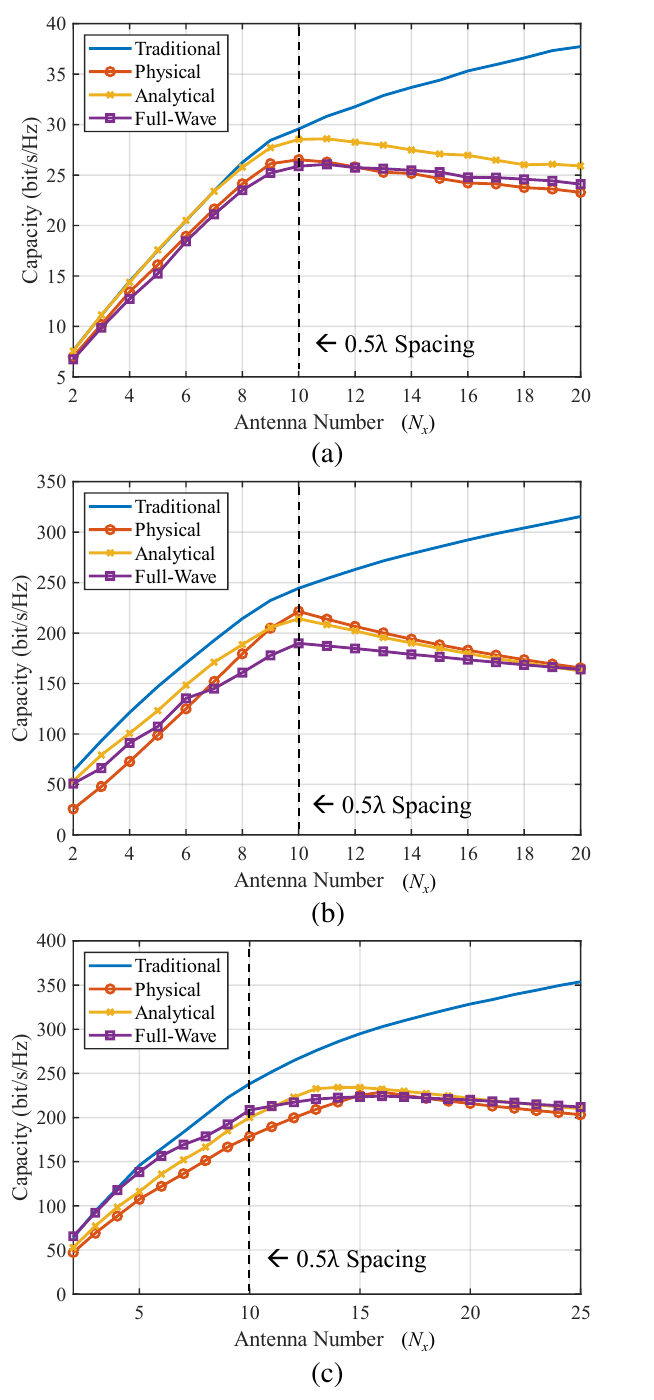}
	\caption{Far-field holographic MIMO capacities using different channel normalization methods in ergodic cases considering antenna efficiency. (a) Linear array. (b) Planar array. (c) Volumetric array.}
	\label{gain_volumetric}
\end{figure}
\subsection{Near-field capacity}
In the near-field scenario, Fig. 10 illustrates the holographic MIMO capacities for both planar and volumetric arrays. The results show that capacity varies with distance depending on the types of losses considered. Polarization and illumination losses are generally minimal, especially at greater distances. However, beamforming losses can be significant; without near-field beamforming, substantial gain reduction occurs at close range, similar to the behavior observed with traditional aperture antennas \cite{kay1960near}. Different from far-field cases, the volumetric array exhibits only a slight advantage over the planar array at very close distances, where the additional DOF provide benefits. 
\begin{figure}[ht!]
	\centering
	\includegraphics[width=3in]{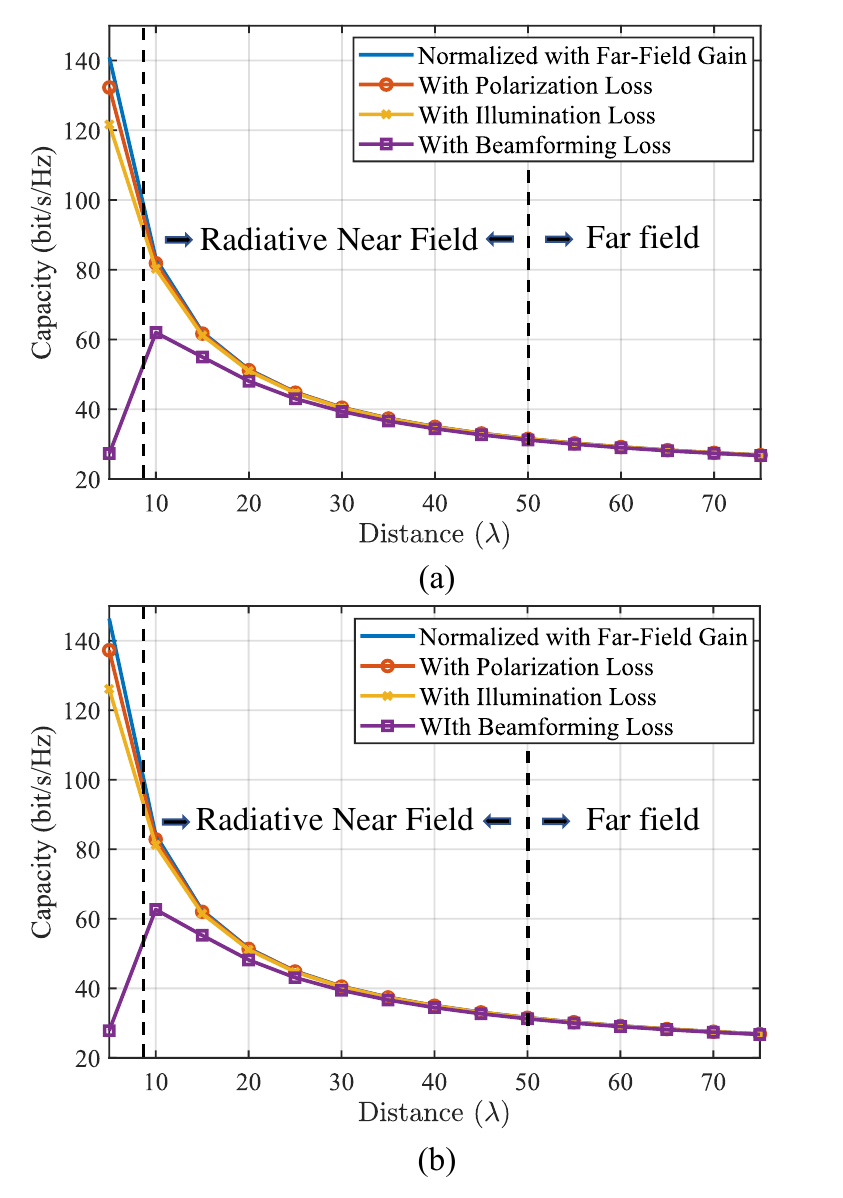}
	\caption{Near-field holographic MIMO capacities with different channel normalization methods. (a) Planar array. (b) Volumetric array.}
	\label{gain_volumetric}
\end{figure}
At larger distances, this advantage diminishes due to the small angular spread, as the effective areas of planar and volumetric arrays are comparable along the vertical direction. While polarization loss is difficult to compensate, illumination and beamforming losses can be effectively managed. Therefore, appropriate normalization methods should be selected based on the specific losses under consideration.
\subsection{Discussion}
Without electromagnetic-based normalization, significant errors may arise in estimating capacity trends and the benefits of dense or volumetric arrays, which could undermine the optimal design of holographic MIMO systems. The proposed normalization methods generally align well with each other, indicating that the choice of method can be adapted to specific needs as follows:
\begin{itemize} 
	\item The full-wave method is the most rigorous method and closely matches practical results but requires substantial computational resources and time. 
	\item The physical method is easier to implement and works well for uniform and regular arrays, although it may be less effective for irregular ones. Additionally, due to complex hardware factors such as specific antenna designs and active antenna impedance, some coefficients in the physical method may need calibration through full-wave simulations. Once these coefficients are determined, the physical method can be applied effectively to specific problems, such as certain volumetric array topologies. 
	\item The analytical method provides a balance between accuracy and computational efficiency, making it a practical compromise between the more demanding full-wave and the simpler physical methods. 
\end{itemize} 

Ultimately, the choice of normalization method should be guided by the specific scenario, the loss factors considered, and the available computational resources.
\section{Conclusion}
This paper addresses the electromagnetic-based normalization of channel matrices for arbitrary array topologies in holographic MIMO communications. It demonstrates that normalization should reflect the realized gains of the antenna array in target directions, which is particularly crucial for holographic arrays due to the coupling and correlation effects inherent in dense or volumetric topologies. The proposed methods for array gains and normalization are validated through analytical, physical, and full-wave methods. Future work will focus on exploring novel holographic array topologies and developing corresponding precoding technologies, including irregular volumetric structures.

\section{Appendix A: Properties of Bessel functions}
Some mathematical properties of Bessel functions are used in this paper. From (V-35) of \cite[Ap. V]{balanis2015antenna}, we have
\begin{equation}\label{62}
J_n(x)=\frac{j^{-n}}{2 \pi} \int_0^{2 \pi} e^{j x \cos \phi} e^{j n \phi} d \phi.
\end{equation}

From (10), \cite[Ch. 13]{luke2014integrals}, we have
\begin{equation}
\begin{aligned}\label{63}
& \int_0^{\pi / 2} J_v(z \sin t) \sin ^{2 \alpha-1} t \cos ^{2 \beta-1} t d t =\\
& \frac{\frac{1}{2}(z / 2)^v \Gamma(\beta) \Gamma(\alpha+v / 2)}{\Gamma(v+1) \Gamma(\alpha+\beta+v / 2)}\times\\ &{}_1{F}_2\left(\left.\begin{array}{l}
a+v / 2 \\
v+1, a+\beta+v / 2
\end{array} \right\rvert\,-z^2 / 4\right) ,
\end{aligned}
\end{equation}
where $\Gamma$ is the Gamma function, ${}_1{F}_2$ is the hypergeometric function with order $(1,2)$.

From (6.677.6), \cite[Ch. 6]{gradshteyn2014table}, we have
\begin{equation}\label{64}
\int_0^a J_0\left(b \sqrt{a^2-x^2}\right) \cos (c x) d x=\frac{\sin \left(a \sqrt{b^2+c^2}\right)}{\sqrt{b^2+c^2}} \quad[b>0].
\end{equation}
\section{Appendix B: Explicit forms of dyadic Green's function}
Explicit forms of electric and magnetic dyadic Green's  function are given here for easily obtaining the near-field results.
\subsection{Electric Green's function}
The electric dyadic Green's function is
\begin{equation}
\bar{\boldsymbol{\mathcal{G}}}\left(R\right)=\left(\begin{array}{ccc}
k^2+\frac{\partial^2}{\partial x^2} & \frac{\partial^2}{\partial x \partial y} & \frac{\partial^2}{\partial x \partial z} \\
\frac{\partial^2}{\partial y \partial x} & k^2+\frac{\partial^2}{\partial y^2} & \frac{\partial^2}{\partial y \partial z} \\
\frac{\partial^2}{\partial z \partial x} & \frac{\partial^2}{\partial z \partial y} & k^2+\frac{\partial^2}{\partial z^2}
\end{array}\right) \frac{e^{-j kR}}{4 \pi k^2R},
\end{equation}
where $R=\sqrt{(x-x^{\prime})^2+(y-y^{\prime})^2+(z-z^{\prime})^2}$, and the $ \bar{\boldsymbol{\mathcal{G}}}$ can be further written as
\begin{equation}
\bar{\boldsymbol{\mathcal{G}}}(R)=\mathcal{G}_1(R) \overline{\mathbf{I}}+\mathcal{G}_2(R) \mathbf{a}_R \mathbf{a}_R,
\end{equation}

\begin{equation}
\mathcal{G}_1(R)=\left(-1-j k R+k^2 R^2\right) \frac{e^{-j k R}}{4 \pi k^2 R^3},
\end{equation}

\begin{equation}
\mathcal{G}_2(R)=\left(3+3 j k R-k^2 R^2\right) \frac{e^{-j k R}}{4 \pi k^2 R^3},
\end{equation}

\begin{equation}
\mathbf{a}_R \mathbf{a}_R=\left[\begin{array}{ccc}
\cos \mathrm{x} \cos\mathrm{x} & \cos \mathrm{y} \cos \mathrm{x} & \cos \mathrm{z} \cos \mathrm{x} \\
\cos \mathrm{x} \cos\mathrm{y} & \cos \mathrm{y} \cos \mathrm{y} & \cos \mathrm{z} \cos \mathrm{y} \\
\cos \mathrm{x} \cos\mathrm{z} & \cos \mathrm{y} \cos \mathrm{z} & \cos \mathrm{z} \cos \mathrm{z}
\end{array}\right],
\end{equation}
where
\begin{equation}
\cos \mathrm{x}=\left(x-x^{\prime}\right) / R, \cos \mathrm{y}=\left(y-y^{\prime}\right) / R, \cos \mathrm{z}=\left(z-z^{\prime}\right) / R.
\end{equation}
\subsection{Magnetic Green's function}
The magnetic dyadic Green's function is
\begin{equation}
\nabla \times  \bar{\boldsymbol{\mathcal{G}}}\left(\mathbf{r}, \mathbf{r}^{\prime}\right)=\left(\begin{array}{ccc}
0 & -\frac{\partial}{\partial z} & \frac{\partial}{\partial y} \\
\frac{\partial}{\partial z} & 0 & -\frac{\partial}{\partial x} \\
-\frac{\partial}{\partial y} & \frac{\partial}{\partial x} & 0
\end{array}\right) \frac{e^{-j kR}}{4 \pi R}.
\end{equation}
The entries of the magnetic Greens function only include one-order derivative, which are
\begin{equation}
\frac{\partial}{\partial x} \left(\frac{e^{-j kR}}{4 \pi R}\right) = -\frac{j kR  \mathrm{e}^{-j k R}\cos \mathrm{x}}{4 \pi R^2}-\frac{R\mathrm{e}^{-j k R} \cos \mathrm{x} }{4\pi R^3},
\end{equation}
\begin{equation}
\frac{\partial}{\partial y} \left(\frac{e^{-j kR}}{4 \pi R}\right) = -\frac{j kR  \mathrm{e}^{-j k R}\cos \mathrm{y}}{4 \pi R^2}-\frac{R \mathrm{e}^{-j k R}\cos \mathrm{y} }{4\pi R^3},
\end{equation}
\begin{equation}
\frac{\partial}{\partial z} \left(\frac{e^{-j kR}}{4 \pi R}\right) = -\frac{j kR  \mathrm{e}^{-j k R}\cos \mathrm{z}}{4 \pi R^2}-\frac{R  \mathrm{e}^{-j k R}\cos \mathrm{z}}{4\pi R^3},
\end{equation}
the definitions of $\cos \mathrm{x},\cos \mathrm{y},\cos \mathrm{z}$ are the same as that in electric dyadic Green's function.

	\bibliographystyle{unsrt}  
\bibliography{Bibliography}

\begin{thebibliography}{10}

\bibitem{shannon}
Claude~E Shannon.
\newblock A mathematical theory of communication.
\newblock {\em Bell system technical journal}, 27(3):379--423, 1948.

\bibitem{telatar1999capacity}
Emre Telatar.
\newblock Capacity of multi-antenna {Gaussian} channels.
\newblock {\em Eur. Trans. Telecomm.}, 10(6):585--595, 1999.

\bibitem{TL2014}
Erik~G. Larsson, Ove Edfors, Fredrik Tufvesson, and Thomas~L. Marzetta.
\newblock Massive {MIMO} for next generation wireless systems.
\newblock {\em IEEE Commun. Mag.}, 52(2):186--195, 2014.

\bibitem{bjornson2019massive}
Emil Bj{\"o}rnson, Luca Sanguinetti, Henk Wymeersch, Jakob Hoydis, and Thomas~L
  Marzetta.
\newblock Massive {MIMO} is a reality—what is next?: Five promising research
  directions for antenna arrays.
\newblock {\em Digit. Signal Prog.}, 94:3--20, 2019.

\bibitem{Huang2020}
Chongwen Huang, Sha Hu, George~C. Alexandropoulos, Alessio Zappone, Chau Yuen,
  Rui Zhang, Marco~Di Renzo, and Merouane Debbah.
\newblock Holographic {MIMO} surfaces for 6{G} wireless networks:
  Opportunities, challenges, and trends.
\newblock {\em IEEE Wirel. Commun.}, 27(5):118--125, 2020.

\bibitem{Pizzo2020}
Andrea Pizzo, Thomas~L. Marzetta, and Luca Sanguinetti.
\newblock Spatially-stationary model for holographic {MIMO} small-scale fading.
\newblock {\em IEEE J. Sel. Areas Commun.}, 38(9):1964--1979, 2020.

\bibitem{lu2021communicating}
Haiquan Lu and Yong Zeng.
\newblock Communicating with extremely large-scale array/surface: Unified
  modeling and performance analysis.
\newblock {\em IEEE Trans. Wirel. Commun.}, 21(6):4039--4053, 2021.

\bibitem{wang2024tutorial}
Zhe Wang, Jiayi Zhang, Hongyang Du, Dusit Niyato, Shuguang Cui, Bo~Ai,
  M{\'e}rouane Debbah, Khaled~B Letaief, and H~Vincent Poor.
\newblock A tutorial on extremely large-scale {MIMO} for 6{G}: Fundamentals,
  signal processing, and applications.
\newblock {\em IEEE Commun. Surv. Tutor.}, 2024.

\bibitem{wei2023tri}
Li~Wei, Chongwen Huang, George~C Alexandropoulos, Zhaohui Yang, Jun Yang,
  EI~Wei, Zhaoyang Zhang, M{\'e}rouane Debbah, and Chau Yuen.
\newblock Tri-polarized holographic {MIMO} surfaces for near-field
  communications: Channel modeling and precoding design.
\newblock {\em IEEE Trans. Wirel. Commun.}, 22(12):8828--8842, 2023.

\bibitem{ji2024electromagnetic}
Ran Ji, Chongwen Huang, Xiaoming Chen, EI~Wei, Linglong Dai, Jiguang He,
  Zhaoyang Zhang, Chau Yuen, and M{\'e}rouane Debbah.
\newblock Electromagnetic hybrid beamforming for holographic {MIMO}
  communications.
\newblock {\em IEEE Trans. Wirel. Commun.}, 2024.

\bibitem{gong2023holographic}
Tierui Gong, Panagiotis Gavriilidis, Ran Ji, Chongwen Huang, George~C
  Alexandropoulos, Li~Wei, Zhaoyang Zhang, M{\'e}rouane Debbah, H~Vincent Poor,
  and Chau Yuen.
\newblock Holographic {MIMO} communications: Theoretical foundations, enabling
  technologies, and future directions.
\newblock {\em IEEE Commun. Surv. Tutor.}, 2023.

\bibitem{Dardari2021}
Davide Dardari and Nicolò Decarli.
\newblock Holographic communication using intelligent surfaces.
\newblock {\em IEEE Commun. Mag.}, 59(6):35--41, 2021.

\bibitem{An2023}
Jiancheng An, Chao Xu, Derrick Wing~Kwan Ng, George~C. Alexandropoulos,
  Chongwen Huang, Chau Yuen, and Lajos Hanzo.
\newblock Stacked intelligent metasurfaces for efficient holographic {MIMO}
  communications in 6g.
\newblock {\em IEEE J. Sel. Areas Commun.}, 41(8):2380--2396, 2023.

\bibitem{franceschetti_2017}
Massimo Franceschetti.
\newblock {\em Wave Theory of Information}.
\newblock Cambridge University Press, 2017.

\bibitem{yuan2024breaking}
Shuai S.~A. Yuan, Jie Wu, Hongjing Xu, Tengjiao Wang, Da~Li, Xiaoming Chen,
  Chongwen Huang, Sheng Sun, Shilie Zheng, Xianmin Zhang, et~al.
\newblock Breaking the degrees-of-freedom limit of holographic {MIMO}
  communications: A 3-{D} antenna array topology.
\newblock {\em IEEE Trans. Veh. Technol.}, 73(8):11276--11288, 2024.

\bibitem{gustafsson2024shadow}
Mats Gustafsson.
\newblock Shadow area and degrees-of-freedom for free-space communication.
\newblock {\em arXiv preprint arXiv:2407.21122}, 2024.

\bibitem{gao2015massive}
Xiang Gao, Ove Edfors, Fredrik Rusek, and Fredrik Tufvesson.
\newblock Massive {MIMO} performance evaluation based on measured propagation
  data.
\newblock {\em IEEE Trans. Wirel. Commun.}, 14(7):3899--3911, 2015.

\bibitem{wallace2003experimental}
Jon~W Wallace, Michael~A Jensen, A~Lee Swindlehurst, and Brian~D Jeffs.
\newblock Experimental characterization of the {MIMO} wireless channel: Data
  acquisition and analysis.
\newblock {\em IEEE Trans. Wirel. Commun.}, 2(2):335--343, 2003.

\bibitem{mabrouk2013feasibility}
Ismail~Ben Mabrouk, Julien Hautcoeur, Larbi Talbi, Mourad Nedil, and Khelifa
  Hettak.
\newblock Feasibility of a millimeter-wave {MIMO} system for short-range
  wireless communications in an underground gold mine.
\newblock {\em IEEE Trans. Antennas Propag.}, 61(8):4296--4305, 2013.

\bibitem{martinez2018experimental}
{\`A}lex~Oliveras Mart{\'\i}nez, Jesper~{\O}dum Nielsen, Elisabeth De~Carvalho,
  and Petar Popovski.
\newblock An experimental study of massive {MIMO} properties in {5G} scenarios.
\newblock {\em IEEE Trans. Antennas Propag.}, 66(12):7206--7215, 2018.

\bibitem{sadeghi2017large}
Meysam Sadeghi, Luca Sanguinetti, Romain Couillet, and Chau Yuen.
\newblock Large system analysis of power normalization techniques in massive
  {MIMO}.
\newblock {\em IEEE Trans. Veh. Technol.}, 66(10):9005--9017, 2017.

\bibitem{Caire2015}
Yeon-Geun Lim, Chan-Byoung Chae, and Giuseppe Caire.
\newblock Performance analysis of massive {MIMO} for cell-boundary users.
\newblock {\em IEEE Trans. Wirel. Commun.}, 14(12):6827--6842, 2015.

\bibitem{Loyka2009}
Sergey Loyka and Georgy Levin.
\newblock On physically-based normalization of {MIMO} channel matrices.
\newblock {\em IEEE Trans. Wirel. Commun.}, 8(3):1107--1112, 2009.

\bibitem{d2024holographic}
Antonio~Alberto D’Amico and Luca Sanguinetti.
\newblock Holographic {MIMO} communications: What is the benefit of closely
  spaced antennas?
\newblock {\em IEEE Trans. Wirel. Commun.}, 2024.

\bibitem{burr2003capacity}
Alister~G Burr.
\newblock Capacity bounds and estimates for the finite scatterers {MIMO}
  wireless channel.
\newblock {\em IEEE J. Sel. Areas Commun.}, 21(5):812--818, 2003.

\bibitem{zhu2024mimo}
Jieao Zhu, Vincent~YF Tan, and Linglong Dai.
\newblock {MIMO} capacity analysis and channel estimation for electromagnetic
  information theory.
\newblock {\em arXiv preprint arXiv:2406.04881}, 2024.

\bibitem{coldrey2008modeling}
Mikael Coldrey.
\newblock Modeling and capacity of polarized {MIMO} channels.
\newblock In {\em VTC Spring 2008-IEEE Vehicular Technology Conference}, pages
  440--444. IEEE, 2008.

\bibitem{qian2024including}
Jun Qian, Shanpu Shen, and Ross Murch.
\newblock Including antenna effects into capacity formulations of line-of-sight
  {MIMO} channels.
\newblock {\em IEEE Wirel. Commun. Lett.}, 2024.

\bibitem{tse2005fundamentals}
David Tse and Pramod Viswanath.
\newblock {\em Fundamentals of wireless communication}.
\newblock 2005.

\bibitem{miller2019waves}
David~AB Miller.
\newblock Waves, modes, communications, and optics: a tutorial.
\newblock {\em Adv. Opt. Photonics}, 11(3):679--825, 2019.

\bibitem{ShuaiPra}
Shuai S.~A. Yuan, Jie Wu, Menglin L.~N. Chen, Zhihao Lan, Liang Zhang, Sheng
  Sun, Zhixiang Huang, Xiaoming Chen, Shilie Zheng, Li~Jun Jiang, Xianmin
  Zhang, and Wei E.~I. Sha.
\newblock Approaching the fundamental limit of orbital-angular-momentum
  multiplexing through a hologram metasurface.
\newblock {\em Phys. Rev. Applied}, 16:064042, 2021.

\bibitem{bai2024information}
Xuyang Bai, Shurun Tan, Said Mikki, Erping Li, and Tie-Jun Cui.
\newblock Information-theoretic measures for reconfigurable metasurface-enabled
  direct digital modulation systems: An electromagnetic perspective.
\newblock {\em Prog. Electromagn. Res.}, 179:1--18, 2024.

\bibitem{larsson2003space}
Erik~G Larsson, Petre Stoica, and Girish Ganesan.
\newblock {\em Space-time block coding for wireless communications}.
\newblock Cambridge university press, 2003.

\bibitem{hochwald2000unitary}
Bertrand~M Hochwald and Thomas~L Marzetta.
\newblock Unitary space-time modulation for multiple-antenna communications in
  rayleigh flat fading.
\newblock {\em IEEE Trans. Inf. Theory}, 46(2):543--564, 2000.

\bibitem{oestges2006validity}
Claude Oestges.
\newblock Validity of the kronecker model for {MIMO} correlated channels.
\newblock In {\em 2006 IEEE 63rd Vehicular Technology Conference}, volume~6,
  pages 2818--2822. IEEE, 2006.

\bibitem{xiaoming2013}
Xiaoming Chen, Per-Simon Kildal, Jan Carlsson, and Jian Yang.
\newblock {MRC} diversity and {MIMO} capacity evaluations of multi-port
  antennas using reverberation chamber and anechoic chamber.
\newblock {\em IEEE Trans. Antennas Propag.}, 61(2):917--926, 2013.

\bibitem{chen2018review}
Xiaoming Chen, Shuai Zhang, and Qinlong Li.
\newblock A review of mutual coupling in {MIMO} systems.
\newblock {\em IEEE Access}, 6:24706--24719, 2018.

\bibitem{Hannan1964}
P.~Hannan.
\newblock The element-gain paradox for a phased-array antenna.
\newblock {\em IEEE Trans. Antennas Propag.}, 12(4):423--433, 1964.

\bibitem{Nicola2022}
Nicola Anselmi, Paolo Rocca, Stefan Feuchtinger, Bruno Biscontini,
  Alejandro~Murillo Barrera, and Andrea Massa.
\newblock Optimal capacity-driven design of aperiodic clustered phased arrays
  for multi-user {MIMO} communication systems.
\newblock {\em IEEE Trans. Antennas Propag.}, 70(7):5491--5505, 2022.

\bibitem{tsang1985theory}
Leung Tsang, Jin~Au Kong, and Robert~T Shin.
\newblock Theory of microwave remote sensing.
\newblock 1985.

\bibitem{Shuai2021}
Shuai S.~A. Yuan, Zi~He, Xiaoming Chen, Chongwen Huang, and Wei E.~I. Sha.
\newblock Electromagnetic effective degree of freedom of an {MIMO} system in
  free space.
\newblock {\em IEEE Antennas Wirel. Propag. Lett.}, 21(3):446--450, 2022.

\bibitem{balanis2015antenna}
Constantine~A Balanis.
\newblock {\em Antenna theory: analysis and design}.
\newblock 2015.

\bibitem{costa2018closed}
Bruno~Felipe Costa and Taufik Abr{\~a}o.
\newblock Closed-form directivity expression for arbitrary volumetric antenna
  arrays.
\newblock {\em IEEE Trans. Antennas Propag.}, 66(12):7443--7448, 2018.

\bibitem{das2016generalization}
Sudipta Das, Durbadal Mandal, Sakti~Prasad Ghoshal, and Rajib Kar.
\newblock Generalization of directivity expressions for antenna arrays.
\newblock {\em IEEE Trans. Antennas Propag.}, 65(2):915--919, 2016.

\bibitem{kedar2018wide}
Ashutosh Kedar and LP~Ligthart.
\newblock Wide scanning characteristics of sparse phased array antennas using
  an analytical expression for directivity.
\newblock {\em IEEE Trans. Antennas Propag.}, 67(2):905--914, 2018.

\bibitem{stegen1964gain}
R~Stegen.
\newblock The gain-beamwidth product of an antenna.
\newblock {\em IEEE Trans. Antennas Propag.}, 12(4):505--506, 1964.

\bibitem{nuttall2001approximations}
Albert~H Nuttall and Benjamin~A Cray.
\newblock Approximations to directivity for linear, planar, and volumetric
  apertures and arrays.
\newblock {\em IEEE J. Ocean. Eng.}, 26(3):383--398, 2001.

\bibitem{ji2024}
Ran Ji, Chongwen Huang, Xiaoming Chen, Wei E.~I. Sha, Zhaoyang Zhang, Jun Yang,
  Kun Yang, Chau Yuen, and Mérouane Debbah.
\newblock Exploring {Hannan} limitation for {3D} antenna array.
\newblock {\em arXiv preprint arXiv:2409.01566}, 2024.

\bibitem{ShuaiOJAP}
Shuai S.~A. Yuan, Xiaoming Chen, Chongwen Huang, and Wei E.~I. Sha.
\newblock Effects of mutual coupling on degree of freedom and antenna
  efficiency in holographic {MIMO} communications.
\newblock {\em IEEE Open J. Antennas Propag.}, 4:237--244, 2023.

\bibitem{Kildal2016}
Per-Simon Kildal, Abbas Vosoogh, and Stefano Maci.
\newblock Fundamental directivity limitations of dense array antennas: A
  numerical study using {H}annan’s embedded element efficiency.
\newblock {\em IEEE Antennas Wirel. Propag. Lett.}, 15:766--769, 2016.

\bibitem{kay1960near}
A~Kay.
\newblock Near-field gain of aperture antennas.
\newblock {\em IRE Trans. Antennas Propag.}, 8(6):586--593, 1960.

\bibitem{xiao2021near}
Luyin Xiao, Yongjun Xie, Peiyu Wu, and Junbao Li.
\newblock Near-field gain expression for aperture antenna and its application.
\newblock {\em IEEE Antennas Wirel. Propag. Lett.}, 20(7):1225--1229, 2021.

\bibitem{luke2014integrals}
Yudell~L Luke.
\newblock {\em Integrals of Bessel functions}.
\newblock Courier Corporation, 2014.

\bibitem{gradshteyn2014table}
Izrail~Solomonovich Gradshteyn and Iosif~Moiseevich Ryzhik.
\newblock {\em Table of integrals, series, and products}.
\newblock Academic press, 2014.

\end{thebibliography}

\vfill


\end{document}